\documentclass[lettersize,journal]{IEEEtran}
\usepackage{amsmath,amsfonts}
\usepackage{array}
\usepackage{comment}
\usepackage[caption=false,font=normalsize,labelfont=sf,textfont=sf]{subfig}
\usepackage{textcomp}
\usepackage{stfloats}
\usepackage{url}
\usepackage{verbatim}
\usepackage{graphicx}
\usepackage{cite}
\def\BibTeX{{\rm B\kern-.05em{\sc i\kern-.025em b}\kern-.08em
    T\kern-.1667em\lower.7ex\hbox{E}\kern-.125emX}}

\usepackage{amsthm} 
\newtheorem{definition}{Definition}[section] 

\usepackage[shortlabels,inline]{enumitem}

\usepackage{pifont}

\usepackage{float}

\usepackage{multirow}

\usepackage{booktabs}

\usepackage{color}
\ifodd 1
\newcommand{\rev}[1]{{\color{black}#1}}
\newcommand{\yrev}[1]{{\color{red}#1}}

\begin{document}
\let\WriteBookmarks\relax
\def\floatpagepagefraction{1}
\def\textpagefraction{.001}



\title{Tracing Human Stress from Physiological Signals using UWB Radar}  

\author{Jia Xu,~\IEEEmembership{Member,~IEEE,} Teng Xiao, Pin Lv$^{\ast}$,~\IEEEmembership{Member,~IEEE,} Zhe Chen,~\IEEEmembership{Member,~IEEE,} Chao Cai,~\IEEEmembership{Member,~IEEE,} Yang Zhang, Zehui Xiong
\thanks{Jia Xu, Pin Lv are with Guangzhou University, Guangzhou, China (e-mail: xujia@gzhu.edu.cn, lvpin@@gzhu.edu.cn), and Teng Xiao is with Guangxi University, Nanning, China (e-mail: xiaoteng@st.gxu.edu.cn).
Zhe Chen is with the Intelligent Networking and Computing Research Center, and the School of Computer Science, Fudan University, Shanghai, China (e-mail: zhechen@fudan.edu.cn).
Chao Cai is with Huazhong University of Science and Technology, Wuhan, China (e-mail: chriscai@hust.edu.cn).
Yang Zhang is with Nanjing University of Aeronautics and Astronautics, Nanjing, China (e-mail: yangzhang@nuaa.edu.cn).
Zehui Xiong is with Singapore University of Technology and Design, Singapore (e-mail: zehui\_xiong@sutd.edu.sg). Corresponding author is Pin Lv.

}
}









\maketitle

\begin{abstract}
Stress tracing is an important research domain that supports many applications, such as health care and stress management; and its closest related works are derived from stress detection. However, these existing works cannot well address two important challenges facing stress detection. First, most of these studies involve asking users to wear physiological sensors to detect their stress states, which has a negative impact on the user experience. Second, these studies have failed to effectively utilize multimodal physiological signals, which results in less satisfactory detection results. This paper formally defines the stress tracing problem, which emphasizes the continuous detection of human stress states. A novel deep stress tracing method, named DST, is presented. Note that DST proposes tracing human stress based on physiological signals collected by a noncontact ultra-wideband radar, which is more friendly to users when collecting their physiological signals. In DST, a signal extraction module is carefully designed at first to robustly extract multimodal physiological signals from the raw RF data of the radar, even in the presence of body movement. Afterward, a multimodal fusion module is proposed in DST to ensure that the extracted multimodal physiological signals can be effectively fused and utilized. Extensive experiments are conducted on three real-world datasets, including one self-collected dataset and two publicity datasets. Experimental results show that the proposed DST method significantly outperforms all the baselines in terms of tracing human stress states. On average, DST averagely provides a 6.31\% increase in detection accuracy on all datasets, compared with the best baselines.
\end{abstract}




\section{Introduction}\label{sec1}
\IEEEPARstart{S}tress is a feeling of physical or mental tension or anxiety that people encounter in \rev{the} case of unavoidable conditions or any thought \rev{that} makes \rev{them} restless, angry, and frustrated~\cite{mittal2022can,fang2024ic3m}.
\rev{There} is evidence that stress is one of the major factors \rev{of} various physiological and psychological diseases \cite{fu2020psychological,zhang2023emotion}.
Therefore, it is extremely critical to design effective methods to trace the evolution of \rev{the} human stress state to \rev{help} individuals manage their stress and help \rev{health care} professionals refine their treatments for stress\rev{-}related diseases.

The closest line of work to the stress tracing problem is the stress detection problem\rev{, } which is an emerging and important research area for developing advanced computational systems to detect, model, and recognize mental stress states.
Since mental stress is frequently accompanied by physical and physiological changes, a person's stress can \rev{generally be} detected by monitoring and fusing his \rev{or} her relevant human signals which can be captured by sensor devices~\cite{giannakakis2019review,tang2024merit}. 
\rev{In recent} years\rev{,} many works \rev{on} stress detection \rev{have been} proposed by researchers \rev{and} can be divided into two \rev{main} categories, namely\rev{,} traditional machine learning\rev{-}based methods~\cite{agarwal2023novel, alshorman2022frontal, salankar2021stress, rodriguez2020towards} and deep learning\rev{-}based methods~\cite{tian2022identification, gedam2021review, li2020stress, yu2020passive, masood2019modeling}. 
Compared with \rev{the} traditional machine learning based methods, works proposed based on deep neural networks \rev{have} the advantages of higher performance, effective scaling of data, no feature engineering, and strong adaptability~\cite{yuan2023graph,lin2024split,chen2022new,lin2023pushing,yuan2024satsense,peng2024sums,chen2021octopus}, which makes them the mainstream paradigm for stress detection.
\rev{In addition to applying} deep learning, researchers \rev{have} also \rev{attempted} to utilize more types of human signals to further \rev{improve} the quality of stress detection~\cite{song2021myomonitor, han2017detecting}.
\rev{These} signals either reflect human physical states (i.e., physical signals) or are related to human physiological states (i.e., physiological signals). 
Detecting mental stress via physical signals (such as body postures and facial expressions), however, is deemed to be less reliable since it is relatively easy for a human to control physical signals to hide his \rev{or} her true stress, especially during social interactions~\cite{chen2021pain}. 
Under such circumstance\rev{s}, many works prefer to \rev{use} physiological signals (such as \rev{an electrocardiogram (ECG), heart rate (HR), and respiratory rate (RR)}) to achieve stress detection since physiological signals can detect stress in a modest, inexpensive, convenient, and easy-to-utilize manner. 
\rev{Although} existing stress detection methods based on physiological signals have \rev{achieved} promising results, they have two crucial limitations\rev{:} 
\begin{enumerate}
    \item \textit{\textbf{\rev{Being} unfriendly to users}}. 
    To collect physiological signals, users are mostly required to wear physiological sensors (such as ECG detector\rev{s}, blood pressure monitor\rev{s}, oximeter\rev{s}, and Empatica E4 wrist band\rev{s}~\cite{carreiro2020wearable}). However, the vast majority of users are reluctant to wear such \rev{devices because of} the potential risks of privacy disclosure and skin irritation or \rev{because of} interference \rev{with} their movements~\cite{smets2018into}.
    \item \textit{\textbf{\rev{These methods fail to utilize effectively} multimodal physiological signals.}} 
    It is widely known that different \rev{modalities} of physiological signal\rev{s} (e.g., \rev{HR and RR}) contain different information and thus can be jointly utilized to improve stress detection performance~\cite{baltaci2016stress}. 
    However, current \rev{studies} mostly \rev{directly} utilize multimodal physiological signals, i.e., simply selecting the most robust modal\rev{ity} of physiological signal\rev{s}~\cite{rodriguez2020towards} or \rev{simply} fusing multimodal physiological signals via a \rev{Multilayer Perceptron Neural network (MLP) (or Convolutional Neural Networks (CNN)~\cite{lin2024efficient})}~\cite{li2020stress}, which restricts the improvement \rev{in} their stress detection performance.
\end{enumerate}

\begin{figure*}[htb]%
    \centering	\includegraphics[]{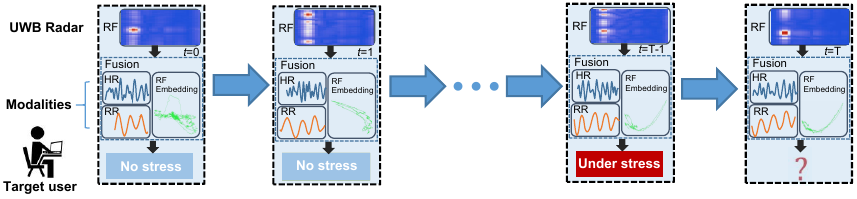}
    \caption{An illustration of \rev{the} stress tracing procedure based on multimodal physiological signals using UWB radar}\label{motivation}
\end{figure*}

Although \rev{a} few works~\cite{ha2021wistress, liang2023mmstress} \rev{that} provide solutions for stress detection based on wireless signals and successfully improve user experience \rev{have recently been proposed}, these works consistently fail to \rev{utilize effectively} multimodal signals to optimize the detection of human stress and thus can only generate less satisfactory detection results. 
\rev{In} this paper, we propose a novel deep learning\rev{-}based method for tracing human stress based on physiological signals called \textit{\textbf{Deep Stress Tracing (DST)}}. \rev{In contrast to} most related works, DST is designed to detect the stress states of a target user in a continuous manner. As shown in Figure \ref{motivation}, DST takes a target user's multimodal physiological signals (i.e., HR and RR)\rev{,} which are derived by decomposing the RF data collected via an ultra-wideband (UWB) radar\rev{,} which is a \rev{contactless} sensor widely used in wireless sensing~\cite{zhang2022quantifying, xu2023ske, lv2023ocro} as the inputs and can continuously trace the change in the user's stress state (i.e., no stress or under stress).
Using UWB radar is the key to address\rev{ing} the first limitation since UWB radar can collect a target user's HR and RR signals in a \rev{noncontact} manner. \rev{Moreover}, UWB radar \rev{has} many advantages in handling stress detection tasks. For example, it has low energy consumption and higher compliance, is not affected by lighting or other physical obstacles, is less susceptible to noise, has fewer privacy issues, and does not have complex equipment deployment~\cite{yang2021overview}. 
 
To \rev{address} the second limitation, DST designs a deep neural architecture \rev{that} optimizes the extraction and utilization of multimodal physiological signals for effective stress tracing. \rev{Specifically}, following the idea in \cite{xie2022deepvs}, 
a \textbf{\textit{Signal Extraction Module} (SEM)} is designed to optimize the extraction of HR and RR signals from RF data. \rev{The} SEM is very robust. On one hand, \rev{the} SEM can still extract high-quality HR and RR signals when \rev{the} HR and RR signals interfere with each other. On the other hand, unintentional body movements\rev{,} which \rev{have a} negative impact on the parameter estimation of physiological movements (i.e., heartbeat and breathing)\rev{,} are \rev{adequately} handled by \rev{the} SEM. 
Then, to better fuse and utilize the multimodal physiological signals (i.e., HR and RR) extracted by \rev{the} SEM, a \textbf{\textit{Multimodal Fusion Module} (MFM)} is presented in DST. \rev{Specifically}, \rev{the} MFM \rev{utilizes} a modular and flexible approach to effectively fuse three types of feature stream\rev{s}, i.e., HR signal\rev{s}, RR signal\rev{s}, and original RF data\rev{,} which contain rich information \rev{such as} head movements and body postures of the target user.
Moreover, to ensure \rev{that} the physiological changes detected at the time of measurement are caused by stress, in this study, we define\rev{d} a stress induction protocol and use\rev{d} the Depression Anxiety Stress Scale (DASS)~\cite{parkitny2010depression} to verify the effectiveness of the protocol.
The main contributions of this paper are summarized as follows:
\begin{itemize}[itemsep=0pt]
\item We formally define the problem of stress tracing\rev{,} which is very important \rev{for} supporting many practical applications, such as health and stress management. 
\item We propose a novel stress tracing method, called Deep Stress Tracing (DST), which effectively traces human stress from physiological signals gather\rev{ed} by an contactless UWB radar. Compared with most related methods, DST is more friendly to users since users are not required to wear any physiological sensors during the stress monitoring procedure. Moreover, DST better exploits multimodal physiological signals and thus improve\rev{s} the detection accuracy compared with related methods built based on contactless radars.
\item We design \rev{several} modules in DST, namely\rev{, the} Signal Extraction Module (SEM) and Multimodal Fusion Module (MFM), to \rev{ensure that} the multimodal physiological signals of a target user are effectively extracted, fused, and utilized to optimize the stress tracing tasks. 
\item The effectiveness of the proposed DST method for stress tracing \rev{is} carefully evaluated using three real-world datasets. \rev{The experimental} results verify the superiority of DST compared with all baselines and the good generalization of DST in handling different multimodal datasets.
\end{itemize}

The \rev{remainder} of the paper is organized as follows \rev{:} Section \ref{sec:rel} analyzes related works. Section \ref{sec:pre} provides preliminaries, where the stress tracing problem is formally defined. Section \ref{sec:method} elaborates the implementation of the proposed Deep Stress Tracing (DST) method. Section \ref{sec:exp} discusses the experimental settings and results. Section \ref{sec:discussion} is dedicated to discuss\rev{ing the} research findings and future work.

\section{Related Works}
\label{sec:rel}
The closest line of work to the stress tracing problem \rev{investigated} in this paper is the stress detection problem \rev{, which} only focuses on predicting a target user's stress state at once and \rev{disregards} the continuity of the target user's stress states. In this section, we review related works about stress detection in terms of the following three aspects.

\subsection{Stress Detection Based on Physical Signals}
Some early studies \rev{proposed detecting} human stress based on physical signals using various machine learning methods. For example, using a set of body language features automatically extracted from visual cues \rev{collected} by a camera as the inputs, Aigrain et al. \cite{aigrain2015person} \rev{applied a} support vector machine (SVM) to perform stress detection.
As another case, a stress detection framework is proposed in~\cite{giannakakis2017stress}, which digests video recording\rev{s} of human facial cues and utilizes five machine learning classification algorithms to \rev{achieve} satisfactory detection accuracy. While the study in~\cite{giannakakis2017stress} \rev{was} also built \rev{on} multimodal features, such as eye-related events, oral activity, and head movements, only the most robust feature \rev{was} selected \rev{for} detection, while the information contained in other multimodal features \rev{was disregarded}.
Baltaci and Gokcay \cite{baltaci2016stress} fully emphasize\rev{d} the importance of fusing multimodal features for stress detection. However, \rev{in} \cite{baltaci2016stress}\rev{ , the authors} \rev{fused} pupil dilation and facial thermal change features based on \rev{a} simple concatenation operation.

Recent works from various domains have demonstrated the power of deep learning in improving the accuracy of classification tasks. Thus, many physical signal\rev{-}based \rev{studies have proposed} stress detection methods based on deep learning techniques. \rev{For} example, Zhang et al.~\cite{zhang2019detecting} present\rev{ed} a connected convolutional network to predict human stress, which is trained via three facial expressions associated with stress, i.e., anger, fear, and sadness. 
As another example, in~\cite{ghosh2022classification}, a new CNN\rev{-}based stress detection architecture is designed, which is trained based on the collected computer records, facial expressions, body postures, and other data of different subjects collected by the camera and wearable devices. 
Although the \rev{authors in}~\cite{zhang2019detecting,ghosh2022classification} propose to \rev{utilize} multimodal data, they \rev{directly} connect the multimodal data in the data preprocessing phase, which will result in the formation of high-dimensional feature spaces without proper fusion and may lead to \rev{greater} computational and storage resources \rev{being needed}.

\subsection{Stress Detection Based on Physiological Signals}
Previous \rev{studies} have shown that analyzing physiological signals is a reliable mean\rev{s} of detecting human stress~\cite{chen2021pain,gedam2021review}. \rev{Many} good stress detection \rev{methods have been} proposed based on the collection and analysis of physiological signals. 

Mozos et al.~\cite{mozos2017stress} propose to combine two sensor systems that separately capture physiological and social responses to detect stress of people in social situations, and the discriminant ability of each sensor system for stress detection is carefully evaluated in the study.
Rodríguez-Arce et al.~\cite{rodriguez2020towards} use\rev{d} Arduino boards and low-cost sensors to predict \rev{the} stress of students in an academic environment. \rev{Although} the detection in this study relies on \rev{the} respiratory rate, heart rate, skin temperature, and blood oxygen \rev{data of students}, it has two limitations: 1) only the most robust physiological feature is \rev{ultimately} used \rev{for} detection \rev{and} 2) inconvenient wearable devices are used to collect students' physiological signals.
Halim and Rehan~\cite{halim2020identification} present\rev{ed} a machine learning approach to identify driving-induced stress patterns, \rev{in which} ongoing brain activity is logged as \rev{an} EEG signal to \rev{determine} the link between brain dynamics and stress states.
Salankar et al.~\cite{salankar2021stress} present\rev{ed} an effective method using MLP and SVM to identify stress markers via the frontal, temporal, central, and occipital lobes. \rev{Specifically}, a variational mode decomposition strategy is used to preprocess multimodal physiological signals, which \rev{slightly} ensures the quality of the extracted physiological signals.
AlShorman et al.~\cite{alshorman2022frontal} \rev{used} frontal EEG spectral analysis to detect stress, where the fast Fourier transform \rev{(FFT)} \rev{was} applied in the feature extraction stage\rev{,} while SVM and Naive Bayes classifiers \rev{were} chosen for stress detection. This approach does not use raw \rev{or} unprocessed data, which allows for more accurate utilization of physiological signal features and\rev{,} as a result\rev{,} guarantees the detection performance of its model.

The idea of deep learning \rev{has} also recently \rev{been} used to enhance stress detection performance based on physiological signals.
First, Masood and Alghamdi~\cite{masood2019modeling} \rev{proposed employing} two wearable devices to monitor people's physiological signals (e.g., heart rate and breathing) and then apply\rev{ing} multiple convolutional layers and recurrent neural networks (RNN\rev{s}) with \rev{long short-term memory (LSTM)} layers \rev{to detect} their stress states. Although this study extracts the brain features from the neural signals to enrich its \rev{modality} information, those multimodal signals are simply concatenated as model inputs, which \rev{disregards} the correlation \rev{among} the features, further causing the features of some modes to contribute less to the final result or to be obscured after the concatenation.
Yu and Sano~\cite{yu2020passive} \rev{built} a stress detection model trained by passive perception data from wearable sensors, mobile phones, and weather APIs. \rev{When} abundant modalities \rev{serve} as input, the model itself is a combination of CNN and LSTM model\rev{s}, which \rev{do} not proper\rev{ly} \rev{employ} multimodal data and may \rev{cause} the introduction of irrelevant or redundant information, even amplifying the effect of noise and affecting the robustness and performance of the model. 
In~\cite{li2020stress}, Li and Liu propose\rev{d} a CNN\rev{-}based solution that analyzes physiological data collected from a sensor worn on the chest of a person to detect his \rev{or} her stress states. This solution also combines multimodal features by concatenating them. 
The work presented in~\cite{de2022mostress} \rev{presented} a new stress model, which relies on a sequential neural network model to predict human stress. The proposed model preprocesses physiological data collected by wearable devices through the RNN \rev{pipeline}. 

\begin{table*}
\begin{center}
\caption{Differences between the proposed DST method and related human stress detection methods}
\label{comparison-with-rel}
\begin{tabular}{|c|ccc|cccc|}
\hline
\multirow{2}{*}{Methods}
    & \multicolumn{3}{c|}{Relying on what to detect human stress} 
    & \multicolumn{4}{c|}{How to detect human stress}                      
    \\ \cline{2-8} 
        & \multicolumn{1}{c|}
            {\begin{tabular}[c]{@{}c@{}}Physiological\\ signals\end{tabular}}
        & \multicolumn{1}{c|}
            {\begin{tabular}[c]{@{}c@{}}Non-contact\\ sensor\end{tabular}}
        & \multicolumn{1}{c|}
            {\begin{tabular}[c]{@{}c@{}}Multimodal\\data\end{tabular}}
        & \multicolumn{1}{c|}
            {\begin{tabular}[c]{@{}c@{}}Continuous\\tracing\end{tabular}}
        & \multicolumn{1}{c|}
            {\begin{tabular}[c]{@{}c@{}}Feature\\fusion\end{tabular}}
        & \multicolumn{1}{c|}
            {\begin{tabular}[c]{@{}c@{}}Information exchange \\among multimodal data\end{tabular}}
        & \begin{tabular}[c]{@{}c@{}}Deep\\learning\end{tabular}
    \\ \hline      

    \multicolumn{1}{|c|}{SD\_HCI~\cite{baltaci2016stress}}  
    & \multicolumn{1}{c|}{\ding{55}} & \multicolumn{1}{c|}{\ding{55}} & \ding{51}   
    & \multicolumn{1}{c|}{\ding{55}} & \multicolumn{1}{c|}{\ding{51}} & \multicolumn{1}{c|}{\ding{55}} & \ding{51} \\ \hline

    \multicolumn{1}{|c|}{SD\_FC~\cite{giannakakis2017stress}}  
    & \multicolumn{1}{c|}{\ding{55}} & \multicolumn{1}{c|}{\ding{51}} & \ding{55}   
    & \multicolumn{1}{c|}{\ding{55}} & \multicolumn{1}{c|}{\ding{55}} & \multicolumn{1}{c|}{\ding{55}} & \ding{55} \\ \hline

    \multicolumn{1}{|c|}{SD\_WPSS~\cite{mozos2017stress}}
    & \multicolumn{1}{c|}{\ding{51}} & \multicolumn{1}{c|}{\ding{55}} & \ding{51} 
    & \multicolumn{1}{c|}{\ding{55}} & \multicolumn{1}{c|}{\ding{55}} & \multicolumn{1}{c|}{\ding{55}} & \ding{55} \\ \hline

    \multicolumn{1}{|c|}{PSNS~\cite{masood2019modeling}}
    & \multicolumn{1}{c|}{\ding{51}} & \multicolumn{1}{c|}{\ding{55}} & \ding{51} 
    & \multicolumn{1}{c|}{\ding{55}} & \multicolumn{1}{c|}{\ding{55}} & \multicolumn{1}{c|}{\ding{55}} & \ding{51} \\ \hline

    \multicolumn{1}{|c|}{SD\_AE~\cite{rodriguez2020towards}}
    & \multicolumn{1}{c|}{\ding{51}} & \multicolumn{1}{c|}{\ding{55}}  & \ding{51} 
    & \multicolumn{1}{c|}{\ding{55}} & \multicolumn{1}{c|}{\ding{55}} & \multicolumn{1}{c|}{\ding{55}} & \ding{55} \\ \hline

    \multicolumn{1}{|c|}{DeepCNN~\cite{li2020stress}}
    & \multicolumn{1}{c|}{\ding{51}} & \multicolumn{1}{c|}{\ding{55}} & \ding{51}
    & \multicolumn{1}{c|}{\ding{55}} & \multicolumn{1}{c|}{\ding{51}} & \multicolumn{1}{c|}{\ding{55}} & \ding{51} \\ \hline

    \multicolumn{1}{|c|}{RTEEG~\cite{alshorman2022frontal}}
    & \multicolumn{1}{c|}{\ding{51}} & \multicolumn{1}{c|}{\ding{55}} & \ding{55}
    & \multicolumn{1}{c|}{\ding{55}} & \multicolumn{1}{c|}{\ding{55}} & \multicolumn{1}{c|}{\ding{55}} & \ding{55} \\ \hline
    
    \multicolumn{1}{|c|}{Mostress~\cite{de2022mostress}}
    & \multicolumn{1}{c|}{\ding{51}} & \multicolumn{1}{c|}{\ding{55}} & \ding{55}
    & \multicolumn{1}{c|}{\ding{55}} & \multicolumn{1}{c|}{\ding{51}} & \multicolumn{1}{c|}{\ding{55}} & \ding{51} \\ \hline

    \multicolumn{1}{|c|}{IEBDNN~\cite{ghosh2022classification}} 
    & \multicolumn{1}{c|}{\ding{51}} & \multicolumn{1}{c|}{\ding{55}} & \ding{55}
    & \multicolumn{1}{c|}{\ding{55}} & \multicolumn{1}{c|}{\ding{51}} & \multicolumn{1}{c|}{\ding{55}} & \ding{51} \\ \hline

    \multicolumn{1}{|c|}{WiStress~\cite{ha2021wistress}} 
    & \multicolumn{1}{c|}{\ding{51}} & \multicolumn{1}{c|}{\ding{51}} & \ding{55}
    & \multicolumn{1}{c|}{\ding{55}} & \multicolumn{1}{c|}{\ding{51}} & \multicolumn{1}{c|}{\ding{55}} & \ding{51} \\ \hline

    \multicolumn{1}{|c|}{mmStress~\cite{liang2023mmstress}} 
    & \multicolumn{1}{c|}{\ding{55}} & \multicolumn{1}{c|}{\ding{51}} & \ding{51}
    & \multicolumn{1}{c|}{\ding{51}} & \multicolumn{1}{c|}{\ding{51}} & \multicolumn{1}{c|}{\ding{55}} & \ding{51} \\ \hline
    
    \multicolumn{1}{|c|}{\textbf{DST (Ours)}}
    & \multicolumn{1}{c|}{\ding{51}} & \multicolumn{1}{c|}{\ding{51}} & \ding{51}
    & \multicolumn{1}{c|}{\ding{51}} & \multicolumn{1}{c|}{\ding{51}} & \multicolumn{1}{c|}{\ding{51}} & \ding{51} \\ \hline
\end{tabular}
\end{center}
\end{table*}

\rev{Although} existing physiological signal\rev{-}based methods \rev{have shown} effectiveness in detecting human stress, most of them rely on wearable sensor devices to collect physiological signals, which are not user \rev{friendly}. On the other hand, these methods fail to \rev{utilize effectively} the collected multimodal physiological signals \rev{when} designing their detection frameworks, while we argue that correlations between different modalities of physiological signal data cannot be \rev{sufficiently} exploited via a simple fusion operation.

\subsection{Stress Detection Based on Wireless Sensing}
In recent years, wireless sensing technology \rev{has} rapidly \rev{developed} due to its high sensing accuracy and \rev{ability to protect privacy and support} many applications\rev{,} such as hand pose estimation~\cite{xu2023ske}, human activity recognition~\cite{ouyang2022cosmo, lv2023ocro}, 
and autonomous driving~\cite{wei2022mmwave, zhang2020device}. \rev{In recent} years\rev{,} few works \rev{have been} proposed to handle human stress detection tasks via wireless signals. 
In~\cite{ha2021wistress}, WiStress is presented to detect human stress states based on the analysis of physiological signals collected by a contactless mmWave radar.
\rev{Specifically}, WiStress sends mmWave signals to a target user, analyzes the signals reflected by the user to derive his \rev{or} her heartbeat, and \rev{then} infers the user's stress states. Then, mmStress~\cite{liang2023mmstress} \rev{was} proposed to detect human stress on the basis of displacement activities related to human stress (i.e., activities generally displayed by a user under stress, such as walking around, scratching, \rev{and} stomping)\rev{,} which are extracted from mmWave signals. WiStress and mmStress have provided solutions for detecting human stress in a \rev{noncontact} manner. However, these two methods either only employ uni-modal signal\rev{s}~\cite{ha2021wistress} or fail to exchange information among multimodal data and thus cannot effectively fuse and utilize multimodal signals \rev{from} target users~\cite{liang2023mmstress} to optimize stress detection tasks, which \rev{results in unsatisfactory} detection results.


In view of \rev{the} limitations of related works, this paper gather\rev{ing} multimodal physiological signals \rev{via noncontact} UWB radar and \rev{applying} the idea of information exchange~\cite{zamir2021multi} to allow information \rev{from} different modalities to be \rev{more effectively} exchanged, fused, and utilized \rev{. Thus,} our proposal, i.e., the DST method, outperforms existing \rev{methods} and has better generalization in handling different multimodal datasets. \rev{Although} we propose \rev{applying} UWB radar rather than mmWave radar in our problem setting, the proposed DST method can also be applied to execute stress tracing tasks based on mmWave radar after simple adaptation.
Table \ref{comparison-with-rel} compares the research gaps between the proposed DST method and related methods, showing the advantages of DST.

\section{Preliminaries}
\label{sec:pre}
In this section, \rev{first,} we introduce the background of RF-based physiological signal (i.e., HR and RR) sensing with signal modeling. 
Second, we formally define the concept of stress tracing and then elaborate \rev{on} the stress tracing problem using UWB radar\rev{,} which is \rev{analyzed} in this paper.

\setlength{\parindent}{0pt} 
\textbf{Background}. The rationale of RF-based physiological signal sensing is that a tiny shift in the chest wall caused by physiological movements (i.e., heartbeat and breathing) changes the propagation distance of the reflected RF signal and thus changes the phase of the received RF signal from which we can extract physiological signals. To collect periodically changing physiological signals using UWB radar, the radar sends frames at regular intervals and superimposes the received frames to form the channel impulse response (CIR) matrix.
Let $\boldsymbol{R}=\{\boldsymbol{r}_1, \boldsymbol{r}_2, \ldots, \boldsymbol{r}_t, \ldots \}$ be a sequence of a target user’s CIR matrices.  ${\boldsymbol{r}_t}\in R$ is in the form of
$\smash{\boldsymbol{r}_t={[r_{t}^{1},\ldots ,\boldsymbol{r}_{t}^{n}, \ldots, \boldsymbol{r}_{t}^{N}]}^T}$\rev{,} the channel impulse response (CIR) matrix received by a UWB radar, where $t$ and $n$ are the \textit{fast-time} \rev{index} and \textit{slow-time} \rev{index}\rev{,} respectively\rev{,} and $N$ is the number of slow-time frames~\cite{shyu2018detection}. Each row of the CIR matrix (denoted as $r_{t}^{n}$) represents the signal received by the radar after sending one frame of signal\rev{s}, while each column of the CIR matrix denotes the change in time over a certain distance of the reflected signal. By modulating the phase from the received CIR matrices and extracting the frequency component, we \rev{can} extract the HR and RR signals.

\begin{definition}
\textbf{\textit{Stress tracing.}} Given continuous observations of physical or physiological signals that can reflect the stress state of the target user, \rev{the aim of} stress tracing \rev{is} to continuously monitor changes in the user's stress state. The output of a stress tracing task for a target user is a time series of stress state\rev{s} in the form of $S=\{s_1, s_2,\ldots,s_t,\ldots\}$, where $s_t\in\{0,1\}$ denotes the predicted stress state of the user w.r.t. the $t^{th}$ time step and $s_t=1$ indicates \rev{that} the user is under stress\rev{,} while $s_t=0$ signals that the user \rev{is} not \rev{under} stress. According to the definition of stress categories based on the ground truth of the samples in the dataset, the task target can be changed from a two-category prediction task to a multi-category prediction task.
\end{definition}

\begin{definition}
\textbf{\textit{Stress tracing problem using UWB radar.}} Let $\boldsymbol{X}=\{\boldsymbol{x}_1, \boldsymbol{x}_2,\ldots,\boldsymbol{x}_t, \ldots\}$ be a sequence of a target user's physiological signals extracted by a sequence of this target user’s CIR matrices $\boldsymbol{R}=\{\boldsymbol{r}_1, \boldsymbol{r}_2, \ldots, \boldsymbol{r}_t, \ldots \}$ collected via a UWB radar. $\boldsymbol{x}_t\in \boldsymbol{X}$ is in the form of $\boldsymbol{x}_t=\{\boldsymbol{hr}_t,\boldsymbol{rr}_t,\boldsymbol{RF}_t\}$, which contains three modalities, namely\rev{,} $\boldsymbol{hr}_t$ (heart rate signal), $\boldsymbol{rr}_t$ (respiratory signal), and $\boldsymbol{RF}_t$ (RF embedding)\rev{,} all \rev{of which are} extracted from the raw RF signal $\boldsymbol{r}_t$ of the UWB radar at time step $t$. The objective of the stress tracing problem using UWB radar is to continuously predict and output the stress state of the target user for every time step based on $\boldsymbol{X}$, i.e., to derive $S=\{s_1, s_2,\ldots,s_t,\ldots\}$\rev{, where} $s_t\in S$ \rev{is} the predicted stress state for the time step $t$. 
\end{definition}

\section{DST Method}\label{sec:method}
In this section, we outline the architecture of the proposed DST method and then elaborate \rev{on} each component implemented in DST.

\subsection{Overview}
\rev{Overall, DST has a multimodal architecture that digests three signal modalities, i.e., heart rate (HR), respiratory rate (RR), and RF embedding (RF) extracted from the raw RF signals of a UWB radar to trace the stress states of target users. 
Figure \ref{framework} depicts the network architecture of the proposed DST method, which mainly has three modules, namely, \textit{signal extraction module} (\textbf{SEM}), \textit{multimodal fusion module} (\textbf{MFM}), and \textit{stress tracing module} (\textbf{STM}). 
First, SEM is responsible for extracting reliable HR and RR signals from the raw RF data collected by a UWB radar. It is noteworthy that the CNN layer and self-attentive mechanism are applied in SEM to address two challenges facing signal extraction, namely non-linearity of signals~\cite{christabel2023kpca} and unintentional body movements of target users~\cite{xie2023rf}. 
Second, MFM is in charge of computing robust fused representations of three modalities. 
In specific, inspired by the idea of coarse-fine-grained parallel feature fusion proposed in~\cite{peng2017ccl} and \cite{jin2022coarse}, the information from the fine-grained modalities (i.e., HR and RR) are exchanged with the coarse-grained modality (i.e., RF embedding) at different stages via a mechanism of cross connections~\cite{velivckovic2016x} in MFM, so that the semantic gap among different modalities can be narrowed~\cite{kose2021multimodal,bhatti2021attentive} during the computation of the fused representations of three modalities. 
Apparently, the proposed MFM makes DST not only strengthen the relationship between coarse-grained (global) features and fine-grained (local) features, but also compensate for the differences in different modalities, to guarantee information from different modalities can be effectively fused and utilized. 
Last, STM models the temporal dynamics and capture the sequential relationships in the continuous stream of those fused representations of three modalities outputted by MFM, which enables a better learning of the evolution of a target user's stress states within the stream and thereby facilitates the tracing of his or her stress states.
} 
Next, we elaborate \rev{on} the implementation details of every module in the proposed DST method.

\subsection{Signal Extraction Module} \label{SEM}
\begin{figure*}[!t]
\centering
\includegraphics[width=0.95\textwidth]{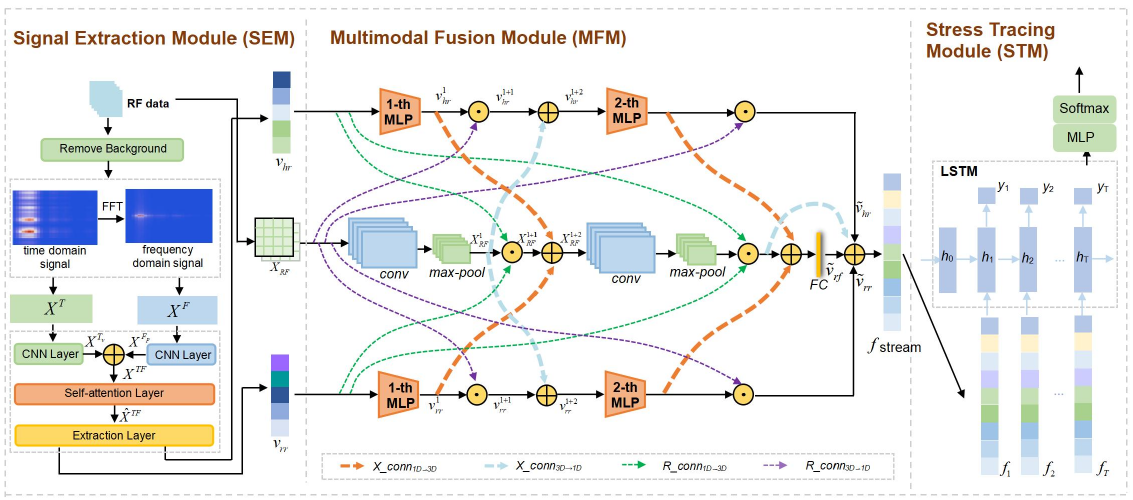}
\caption{\rev{Architecture} of the DST method. \rev{The cross} connection and residual connection in MFM are represented by $X\_conn$ and $R\_conn$, respectively. The former is illustrated \rev{by} a thick dashed line, while the latter is illustrated by a thin dashed line. $\odot $ denotes the element-wise addition operation\rev{,} and $\oplus$ is the concatenation operation.}
\label{framework}
\end{figure*}

The signal extraction module (SEM) targets \rev{the extraction of} reliable HR and RR signals from the raw RF data collected by a UWB radar with the two challenges facing signal extraction, i.e., non-linearity of signals~\cite{christabel2023kpca} and unintentional body movements of the target user~\cite{xie2023rf}.
The implementation details of \rev{the} SEM \rev{are} displayed in Figure \ref{framework}. First, the background noise is removed from the raw RF data by \rev{the} mean difference of the signal technique, which is a commonly used\rev{,} simple but effective signal preprocessing method in the wireless signal processing domain~\cite{li2023simplified, klein2023antenna}. 
Second, a fast Fourier transform (FFT) is applied to convert the time domain signal (denoted by \rev{the} matrix $\boldsymbol{X}^T\in\mathbb{R}^{L\times M}$) of the RF data \rev{in} which background noise has been filtered out into the frequency domain signal (denoted by \rev{the} matrix $\boldsymbol{X}^F\in\mathbb{R}^{L\times N}$). Here, $L$ is the cardinality of continuous time frames in sample data, and $M$\rev{,} $N$ are the number of time samples and the number of spectral samples \rev{respectively} in a time frame.
Then, three stacked layers, i.e., the CNN layer, self-attention layer, and extraction layer, are successively used to process \rev{the} derived time domain input\rev{s} and frequency domain input\rev{s}. In the following \rev{section}, we \rev{separately} explain the implementation of the three layers. 

\subsubsection{CNN Layer}
The CNN layer in \rev{the} SEM adopts the typical two-stream scheme~\cite{bi2023two} to process the derived time domain input $\boldsymbol{X}^T$ and frequency domain input $\boldsymbol{X}^F$ in parallel. 
Note that $\boldsymbol{X}^T$ (or $\boldsymbol{X}^F$) contains a series of continuous time frames of signals in the time domain (or frequency domain). With $\boldsymbol{X}^T$ and $\boldsymbol{X}^F$ as the inputs, \rev{a} one-dimensional CNN is separately applied to $\boldsymbol{X}^T$ and $\boldsymbol{X}^F$ to extract the periodic variations from the time domain $\boldsymbol{X}^T$ (represented as a matrix $\boldsymbol{X}^{T_v}\in \mathbb{R}^{L\times C}$ with $C$ being the number of output channels in the CNN layer) and the condensed spectral peaks from the frequency domain $\boldsymbol{X}^F$ (denoted by \rev{the} matrix $\boldsymbol{X}^{F_p}\in \mathbb{R}^{L\times {{C}'}}$ with ${C}'$ being the number of output channels in the CNN layer), \rev{respectively}. $\boldsymbol{X}^{T_v}$ and $\boldsymbol{X}^{F_p}$ indicate the existence of HR and RR signals in the RF data. 
Then, the concatenation of $\boldsymbol{X}^{T_v}$ and $\boldsymbol{X}^{F_p}$, i.e., $\boldsymbol{X}^{TF}=\boldsymbol{X}^{T_v}\oplus \boldsymbol{X}^{F_p} \in \mathbb{R}^{L\times (C+C')}$, which encodes the information of both \rev{the} time domain and \rev{the} frequency domain, is fed into the following self-\rev{attention} layer for further process\rev{ing}.

\subsubsection{Self-attention Layer}
Note that the physiological HR and RR signals of the target user change over time and display the internal temporal correlation within the observation sequence of HR and RR. In view of this, a multi-head self-\rev{attention} mechanism~\cite{vaswani2017attention} is applied in the attention layer to capture the temporal correlation among the time frames in $\boldsymbol{X}^{TF}$, which is \rev{output} by the previous CNN layer and encodes the information from both time domain and frequency domain. \rev{Specifically}, we apply an $h$-head attention mechanism. The encoded matrix of $\boldsymbol{X}^{TF} \in \mathbb{R}^{L\times (C+C')}$ after the multi-head attention, denoted as $\boldsymbol{\hat{X}^{TF}} \in \mathbb{R}^{L\times (C+C')}$, is computed by: 
\begin{equation}
\begin{aligned}[c]
\begin{array}{ll}
    \boldsymbol{\hat{X}^{TF}}&={MultiHead}\left (\boldsymbol{X}^{TF} \right ) \\
    &=\left ( {head}_1\oplus {head}_2 \ldots \oplus {head}_h \right )\boldsymbol{W}^O, \\
    \text{where} \hspace{0.1em} &{head}_i = \tiny {Attention} \left ( \boldsymbol{X^{TF}}W_i^Q, \boldsymbol{X^{TF}}W_i^K, \boldsymbol{X^{TF}}W_i^V  \right ),
\end{array}
 \label{eq:attention}
\end{aligned}
\end{equation}
where $h$ represents the \rev{number} of heads, $W^O \in \mathbb{R}^{h\times d_k\times (C+C')}$ denotes the weight matrix of \rev{the} output linear transformation with $d_k=\frac{1}{h}(C+C')$, $W_i^Q,W_i^K,W_i^V \in \mathbb{R}^{(C+C') \times d_k}$ are learnable projection matrices for the $i^{th}$ head corresponding to \rev{the} query, key, and value, respectively.

\subsubsection{Extraction Layer}
In the extraction layer, the representations of physiological signals HR and RR are extracted from the outputted matrix of the previous self-attention layer, i.e., $\boldsymbol{\hat{X}^{TF}} \in \mathbb{R}^{L\times (C+C')}$, which encodes the information from \rev{the} time domain and \rev{the} frequency domain. The extraction layer performs an average pool\rev{ing} operation toward $\boldsymbol{\hat{X}^{TF}}$ to aggregate the information of each row in the matrix, which derives an $L$-dimensional aggregation vector, denoted as $\boldsymbol{v}_{tf} \in \mathbb{R}^L$. Then, the extraction layer\rev{,} which is \rev{a MLP} network is used to learn the representations of \rev{the} HR and RR in parallel. \rev{Specifically}, the computation of the representation of HR (denoted as $\boldsymbol{v}_{hr}\in \mathbb{R}^{d_{hr}}$) and the representation of RR (denoted as $\boldsymbol{v}_{rr}\in \mathbb{R}^{d_{rr}}$) follows Equation \ref{eq:hr-rr}.

\begin{subequations}\label{eq:hr-rr}
\begin{align}
\boldsymbol{v}_{hr}&={ReLU} \left ({FC}(\boldsymbol{v}_{tf}) \right )={ReLU} \left (\boldsymbol{v}_{tf}\boldsymbol{W}_{hr}+b_{hr} \right ) \label{eq:hr-}\\
\boldsymbol{v}_{rr}&={ReLU} \left ({FC}(\boldsymbol{v}_{tf}) \right )={ReLU} \left (\boldsymbol{v}_{tf}\boldsymbol{W}_{rr}+b_{rr} \right ), \label{eq:rr-}
\end{align}
\end{subequations}
where $\boldsymbol{W}_{hr}\in \mathbb{R}^{L\times d_{hr}}$ and $\boldsymbol{W}_{rr} \in  \mathbb{R}^{L\times d_{rr}}$ are weight matrices, $b_{hr}$ and $b_{rr}$ are bias terms, and ReLU is a rectified linear unit activation function. 

\rev{In summary}, the proposed signal extraction module (SEM) in the DST method can extract the representations of HR and RR signals in a robust way. \rev{On the} one hand, \rev{the} SEM applies a CNN layer, which enables \rev{learning of} the nonlinear relationship between HR \rev{signals} and RR signals in the time domain (or frequency domain)\rev{. Thus, the} SEM can \rev{adequately} handle the first challenge of non-linearity of signals\rev{,} which is mentioned at the beginning of Section \ref{SEM}. On the other hand, the self-\rev{attention} mechanism used in SEM can capture the temporal correlations among \rev{the} time frames of the received RF data, which is very helpful in decreasing the negative impact \rev{of} the second challenge, i.e., unintentional body movements of the target user. 
 
\subsection{Multimodal Fusion Module} 
\label{MFM}
As shown in Figure \ref{framework}, the multimodal fusion module (MFM) effectively \rev{fuses} three physiological \rev{signal modalities} of the target user by considering the information exchange among these modalities to ensure that the information from all modalities can be fully utilized in the follow-up stress prediction procedure.

The representations of HR (i.e., $\boldsymbol{v}_{hr}\in \mathbb{R}^{d_{hr}}$) and RR ($\boldsymbol{v}_{rr}\in \mathbb{R}^{d_{rr}}$) outputted by the previous SEM become the first two modalities digested by the MFM, while RF embedding is the last modality consumed by the MFM. Here, RF embedding is a complementary feature \rev{that combines} the real part feature \rev{of the raw RF data,} which reflects the signal strength information\rev{,} and \rev{the} imaginary part feature\rev{,} which contains the signal phase information. \rev{Specifically}, the RF embedding is represented as a matrix $\boldsymbol{X}_{RF} \in \mathbb{R}^{L\times M \times 2}$\rev{,} and it \rev{has} two channels\rev{:} a real part matrix $\boldsymbol{X}^r \in \mathbb{R}^{L\times M}$ and an imaginary part matrix $\boldsymbol{X}^i \in \mathbb{R}^{L\times M}$.
We introduce RF embedding as the third processing modality in \rev{the} MFM due to the rich information \rev{that it contains} (i.e., HR, RR, head movement, and body posture of the target user), which is very helpful in improving the performance \rev{of} stress tracing. 

Overall, \rev{the} MFM employs a\rev{n} \textit{MLP×CNN×MLP} structure to learn a high-quality fusion representation of the three modalities of physiological signals by effectively exchanging information between different modalities. 
In particular, cross connections and residual connections are concurrently employed in \rev{the} MFM to implement the information exchange between \rev{the} HR (or RR) stream and \rev{the} RF embedding stream when fusing those three modalities of signals.
Both the cross connections and \rev{the} residual connections can facilitate the learning \rev{of} cross-modal representations in a neural network layer, i.e., they match the shape of multimodal data to pass the information from one modality to another modality in an end-to-end learnable fashion. 
In \rev{the} MFM, both the cross connection and the residual connection are applied multiple times at different processing stages of \rev{the} MFM to ensure adequate information exchange between signal streams of different modalities. Next, we introduce cross connections and residual connections in detail.

\subsubsection {Cross Connection}
As shown \rev{in} Figure \ref{framework}, a cross connection is illustrated by a thick dashed line in \rev{the} MFM, which exchange\rev{s} information \rev{across} different modalities of signals. 
Specifically, cross connections focus on performing the information exchange between the intermediate representations of two signal streams having different modalities at each processing stage of \rev{the} MFM. During an information exchange procedure, the exchanged information from one stream is fused with the intermediate representation of the other stream.
Note that different modalities of signals generally have different dimensions. Such as an HR (or RR) observation is represented as a 1D vector, while an RF embedding is a 3D vector.
Due to the fundamental incompatibility between 1D \rev{data} and 3D data, as shown in Figure \ref{framework}, the intermediate representation of \rev{the} HR (or RR) stream learned by a MLP layer cannot be directly delivered and fused with the intermediate representation of \rev{the} RF embedding stream\rev{,} which is learned by a \{conv, max-pool\} block, and vice versa. 
Therefore, more complicated implementation of cross connection\rev{s} is required to enable information exchange between different modalities of signals in a sensible manner and \rev{to} allow useful interpretations of these transfers. 
Thus, in \rev{the} MFM, we propose two types of cross connection\rev{s}, namely\rev{,} 3D$\to$1D cross connection and 1D$\to $3D cross connection to execute the information exchange between signal streams of HR (or RR) and RF embedding. 

\begin{centering}
\begin{figure}[h]
\centering 
\includegraphics[width=1\linewidth]{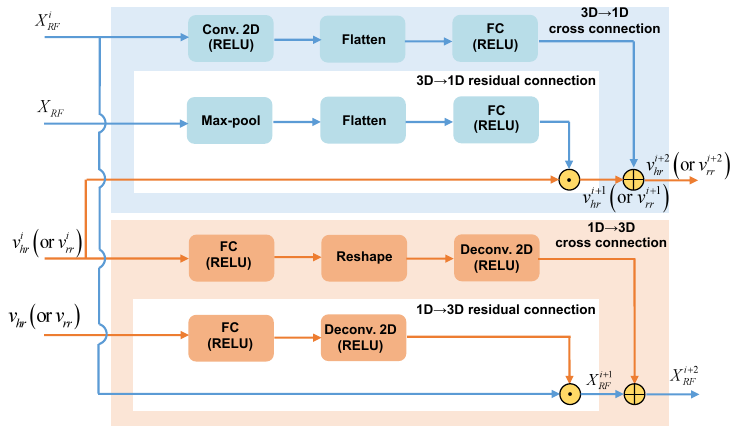}
\caption{\rev{Architecture} of each type of cross connection embedded with a residual connection}
\label{connections}
\end{figure}
\end{centering}

The 3D$\to $1D cross connection (denoted as $X\_conn_{3D\rightarrow1D}$)\rev{,} which is illustrated as a \rev{thick blue} dashed line in Figure \ref{framework}\rev{,} is designed to pass the information contained in the RF embedding stream to the HR (or RR) stream, \rev{whose} structure \rev{is} demonstrated in the upper half of Figure \ref{connections}. As shown \rev{in} the figure, using the output of the $i^{th}$ \{conv, max-pool\} block (denoted as $\boldsymbol{X}_{RF}^i$) in \rev{the} CNN as the input, $X\_conn_{3D\rightarrow1D}$ processes the input \rev{via} a 2D convolutional layer. 
The processing result is then flattened and processed by a fully connected layer, whose result is concatenated with the intermediate representation of HR (or RR) (denoted as $\boldsymbol{v}_{hr}^i$, $\boldsymbol{v}_{rr}^i$), and the result of the 3D$\to $1D cross connection is presented as $\boldsymbol{v}_{hr}^{i+2}$ (or $v_{rr}^{i+2}$).

The 1D$\to $3D cross connection (denoted by $X\_conn_{1D\rightarrow3D}$)\rev{,} which is \rev{represented} by an orange thick dashed line in Figure \ref{framework}\rev{,} is responsible for transferring the information contained in the HR (or RR) stream to the RF embedding stream. 
The implementation details of $X\_conn_{1D\rightarrow3D}$ \rev{are} provided \rev{in} the lower part of Figure \ref{connections}. As shown in the figure, $\boldsymbol{v}_{hr}^i$ (or $\boldsymbol{v}_{rr}^i$) acts as the input of $X\_conn_{1D\rightarrow3D}$. Note that $\boldsymbol{v}_{hr}^i$ (or $\boldsymbol{v}_{rr}^i$) represents the intermediate representation of the HR (or RR) outputted by the $i^{th}$ MLP layer in \rev{the} MFM, which \rev{is} displayed in Figure \ref{framework}. 
$X\_conn_{1D\rightarrow3D}$ processes the input by a fully connected layer, and the processing result is then reshaped and processed by a 2D deconvolution layer \rev{that} converts the 1D representation of the signal to its 3D representation during the training process.
\rev{The} output of the 2D deconvolution layer is concatenated with $\boldsymbol{X}_{RF}^i$, i.e., the intermediate representation of RF embedding outputted by the $i^{th}$ \{conv, max-pool\} block, to obtain the result of the 1D$\to $3D cross connection (presented as $\boldsymbol{X}_{RF}^{i+2}$). 
 
\subsubsection{Residual Connection}
As displayed in Figure \ref{framework}, a residual connection is denoted by \rev{a} thin dashed line in \rev{the} MFM. The residual connection is the process of \rev{directly} adding inputs to the output of the network layer in residual learning to ensure that the information can be \rev{more precisely} passed and updated in the network~\cite{he2016deep}. 
The purpose of introducing the residual connection into \rev{the} MFM is to ensure that the representation of HR, RR, or RF embedding extracted by the signal extraction module (i.e., $\boldsymbol{v}_{hr}, \boldsymbol{v}_{rr}, \boldsymbol{X}_{RF}$) can precisely exchange its information with the intermediate representation of another modality computed by each processing stage in \rev{the} MFM. 
Figure \ref{connections} also illustrates the architecture of \rev{the} two types of residual connection\rev{s} proposed in \rev{the} MFM\rev{:} 1D$\to$3D residual connection (denoted by $R\_conn_{1D\rightarrow3D}$) and 3D$\to$1D residual connection (denoted by $R\_conn_{3D\rightarrow1D}$). 
Specifically, $R\_conn_{3D\rightarrow1D}$ processes the input $\boldsymbol{v}_{rf}$ through a max-pooling layer. The processing result is then flattened and processed by a fully connected layer, whose result is added with the intermediate representation of HR (or RR) (i.e., $\boldsymbol{v}_{hr}^i$, $v_{rr}^i$) by an element-wise addition operation. The addition result is the output of the 3D$\to $1D residual connection, \rev{which is represented} as $\boldsymbol{v}_{hr}^{i+1}$ (or $\boldsymbol{v}_{rr}^{i+1}$).
\rev{Different from} $R\_conn_{3D\rightarrow1D}$, $R\_conn_{1D\rightarrow3D}$ \rev{uses} $\boldsymbol{v}_{hr}$ (or $\boldsymbol{v}_{rr}$) as the input, which is then \rev{successively} processed by a fully connected layer and a 2D deconvolution layer, \rev{the} output of the 2D deconvolution layer is added with the intermediate representation of RF embedding (i.e., $\boldsymbol{X}_{RF}^{i}$) to form the final output result of the 1D$\to$3D residual connection, which is denoted as $\boldsymbol{X}_{RF}^{i+1}$.

As shown in Figure \ref{connections}, a residual connection is straightforwardly embedded into its counterpart cross connection. Such \rev{an} architecture that concurrently applies cross connections and residual connections to achieve information exchange among streams of different modalities provides \rev{the} potential benefit of correcting unwanted impacts that may \rev{be caused} by information exchange. 
For example, to pass information from the HR stream to the RF embedding stream, the 1D$\to$3D cross connection embedded with a residual connection is \rev{employed}, which \rev{uses} $\boldsymbol{v}_{hr}^i$ (i.e., the output of the $i^{th}$ MLP layer that processes HR in \rev{the} MFM) and $\boldsymbol{v}_{hr}$ as the inputs and outputs the new intermediate representation of RF embedding encoding the HR information\rev{,} which is denoted as $\boldsymbol{X}_{RF}^{i+2}$ and computed via Equations \ref{eq:res-conn}-\ref{eq:x-conn}.
\begin{equation}
\begin{aligned}[b]
\boldsymbol{X}_{RF}^{i+1} = \boldsymbol{X}_{RF}^i + \left({{\boldsymbol{W}}_{v_{hr}}}{v_{hr}}+{b_{v_{hr}}} \right)*{{\boldsymbol{K}}_{v_{hr}}}
 \label{eq:res-conn}
\end{aligned}
\end{equation}
\begin{equation}
\begin{aligned}[b]
\boldsymbol{X}_{RF}^{i+2} = \boldsymbol{X}_{RF}^{i+1} \oplus \operatorname{Reshape} \left({{\boldsymbol{W}}_{v_{hr}^i}}{v_{hr}^i}+{b_{v_{hr}^i}} \right)*{\boldsymbol{K}_{v_{hr}^i}} 
,
 \label{eq:x-conn}
\end{aligned}
\end{equation}
where ${\boldsymbol{W}}_{v_{hr}}, {\boldsymbol{W}}_{v_{hr}^i}, {\boldsymbol{K}}_{v_{hr}}$ and ${\boldsymbol{K}}_{v_{hr}^i}$ are learnable weights, ${\boldsymbol{b}}_{v_{hr}}$ and ${\boldsymbol{b}}_{v_{hr}^i}$ are learnable biases.
 
Note that the cross connection embedded with a residual connection is used in each processing stage of \rev{the} MFM to enhance the information exchange between different modalities of signals. As shown in Figure \ref{framework}, in \rev{the} MFM, the final representation of every modality of signal (i.e., ${{\boldsymbol{\tilde{v}}}_{hr}}$, ${{\boldsymbol{\tilde{v}}}_{rr}}$, or ${{\boldsymbol{\tilde{v}}}_{rf}}$)\rev{,} which absorbs the information from other modalities of signals\rev{,} \rev{is} concatenated. The concatenation result $\boldsymbol{f}= {{\boldsymbol{\tilde{v}}}_{rr}} \oplus {{\boldsymbol{\tilde{v}}}_{hr}}\oplus {{\boldsymbol{\tilde{v}}}_{rf}} \in \mathbb{R}^{d_{f}}$ is the learned fusion representation of three modalities of signals which is outputted by \rev{the} MFM and then used by the following stress tracing module (STM) to predict the stress state of the target user.

\subsection{Stress Tracing Module}
Using $\boldsymbol{f}_t$, i.e., the fusion representation of multimodal physiological signals of the target user at a certain time step $t$, as the input, the stress tracing module (STM) is \rev{used to predict the} stress state \rev{of} the target user at the time step.
As shown in Figure \ref{framework}, $\boldsymbol{f}_t$ is fed to an LSTM layer as the $t^{th}$ element of the $\boldsymbol{f}$ stream in \rev{the} STM, and the LSTM layer is responsible for effectively learning the long-term dependencies within the $\boldsymbol{f}$ stream \rev{to} better \rev{predict} the stress states of the target user. \rev{Specifically}, the output of \rev{the} LSTM layer in the form of $\boldsymbol{o}_{t}$ is defined by:
\begin{equation}
\begin{aligned}[b]
\begin{array}{ll}
    \boldsymbol{o}_{t}={f_o} \left (\boldsymbol{h}_t \right ), \\[2ex]
    \text{where} \hspace{0.5em} \boldsymbol{h}_{t}={f_h} \left (\boldsymbol{f}_t, \boldsymbol{h}_{t-1} \right ),
\end{array}
 \label{eq:lstm}
\end{aligned}
\end{equation}

where $\boldsymbol{h}_0\in \{0\}^{d_h}$, and $f_h$, $f_o$ are a state transition function and an output function, respectively.

Then, the output of the LSTM layer $\boldsymbol{o}_{t}$ at time step $t$ is passed to a fully connected layer and a softmax activation in \rev{the} STM to \rev{obtain} $y_t$, which denotes the predicted probability that the target user is under stress at time step $t$. \rev{Specifically}, $y_t$ is computed by:
\begin{equation}
y_t={softmax} \left({{\boldsymbol{W}}_{o_{t}}} \boldsymbol{o}_t + {{\boldsymbol{b}}_{o_{t}}} \right),
 \label{eq:pred}
\end{equation}
where ${\boldsymbol{W}}_{o_{t}}\in \mathbb{R}^{h \times d}$ is the learnable weight matrix, ${\boldsymbol{b}}_{o_{t}}$ is the learnable bias. In the training phase, the average loss is computed by the binary cross\rev{-}entropy\rev{, which is} defined as follows:
\begin{equation}
L=\frac{1}{N}\sum\limits_{n=1}^{N}{\sum\limits_{t=0}^{T^n}{\hat{y}_t^n\log y_{t}^{n}+\left( 1-\hat{y}_t^n \right)}}\log \left( 1-y_t^n \right), 
 \label{eq:loss}
\end{equation}
where $N$ is the cardinality of the $\boldsymbol{f}$ stream and $T^n$ is the latest time step in the $\boldsymbol{f}$ stream. ${y}_t^n$ is the predicted probability that the target user is under stress at time step $t$, and $\hat{y}_t^n \in \{0,1\}$ is the ground-truth label of the stress state of the target user at time step $t$.

\section{Experiments}\label{sec:exp}
\subsection{Stress-inducing Protocol}
\label{subsec:proto}
Determining whether a target user is under stress based on physiological signals is a non-trivial task since we must ensure that the changes \rev{in} measured physiological signals are caused by human stress and not by other factors such as emotions, physical activity, and environmental factors. Inspired by related \rev{studies}~\cite{alshorman2022frontal, li2020stress, masood2019modeling}, we propose \rev{inducing} stress by presenting known stress activities to the target user.
In particular, we design a stress-inducing protocol to ensure \rev{that} the physiological signals used for predicting the stress states of the target user are collected \rev{when} the target user is under stress. 
The stress\rev{-}inducing protocol contains three types of tasks, namely\rev{, the} neutral task, non-stress task, and stress task, which we \rev{describe} below.

\begin{itemize}
\item \textbf{Neutral task.} The neural task is similar to the anticipation period task 
in~\cite{sandulescu2015stress}. In \rev{this} task, a target user is required to \rev{sit} and \rev{answer} simple questions such as "\textit{How are you feeling?}" \rev{and} "\textit{How are you doing today?}". The goal of the task is to allow the target user to rest between different types of tasks.

\item \textbf{Non-stress task.} \rev{In the} non-stress task\rev{,} the target user \rev{is asked} to listen to instrumental music while the user is seated \rev{and} as relaxed as possible.
Sometimes, for certain individual\rev{s}, a task that induces stress may paradoxically be more relaxing than a task without stress. 
To address this issue, the target user is instructed to close their eyes, imagine relaxing scenery, and minimize any thoughts of tension to enhance the state of relaxation. For instance, research~\cite{ogba2019effectiveness} indicates that closing one's eyes while listening to music can alleviate stress.

 
\item \textbf{Stress task.} Previous works~\cite{salai2016stress,sandulescu2015stress} \rev{have shown} that a question-and-answer task with a time limit may induce stress \rev{in} the target user. 
In~\cite{masood2019modeling}, the stress-inducing protocol \rev{used} multiple stress tasks to ensure that participants \rev{felt} stressed while \rev{they were performing} these stress tasks.
In our setting, the target user is asked to \rev{complete} a Stroop color-word interference test (\textbf{stress task A}) and a mental math test (\textbf{stress task B}), which are widely used stress tasks in stress detection\rev{, in a limited amount of time.} \rev{Specifically}, the stress task A requires the target user to select the color of the word while that word expressed literally is printed in \rev{a} different color. When handling stress task A, the target user's stress is caused by the contradiction between verbal perception and visual perception. \rev{In contrast to} stress task A, stress task B induces stress \rev{in} the target user by asking the user to choose the correct option for the computation of a particular value. 
\rev{Note} that the negative feedback (i.e., "\textit{Wrong answer!}" or "\textit{Too slow!}") \rev{was} displayed to the target user \rev{to strengthen} the stress induction in stress task A or B if the answer given by the user \rev{was} incorrect or not submitted within the limited time. Moreover, the remaining time for answering each question is marked with a countdown, and the question-answering interface is displayed with different colors of flashing borders (i.e., green, orange, \rev{and} red) at different time \rev{points} when answering the question to ensure the effectiveness of stress induction. 
To further stimulate the stress of the target user, within a stress task, the user is given either a candy (correct answer) or an electrical stimulation (incorrect answer) according to his \rev{or} her question-answering performance. According to \rev{a} study~\cite{rhudy2000fear}, \rev{expose} to electrical stimulation causes fear, whereas the anticipation of electrical stimulation causes stress. In this study, electrical stimulation \rev{consisted} of a brief pulse with \rev{a} duration of 500 \rev{ms and a current of} less than $5.0$ mA (mild stimulating sensation \rev{without} pain). \rev{Notably,} it is the content and setting of the task rather than the length of the task that causes stress for the target user.
\end{itemize}

\begin{centering}
\begin{figure}[h]
\centering 
\includegraphics[width=1\linewidth]{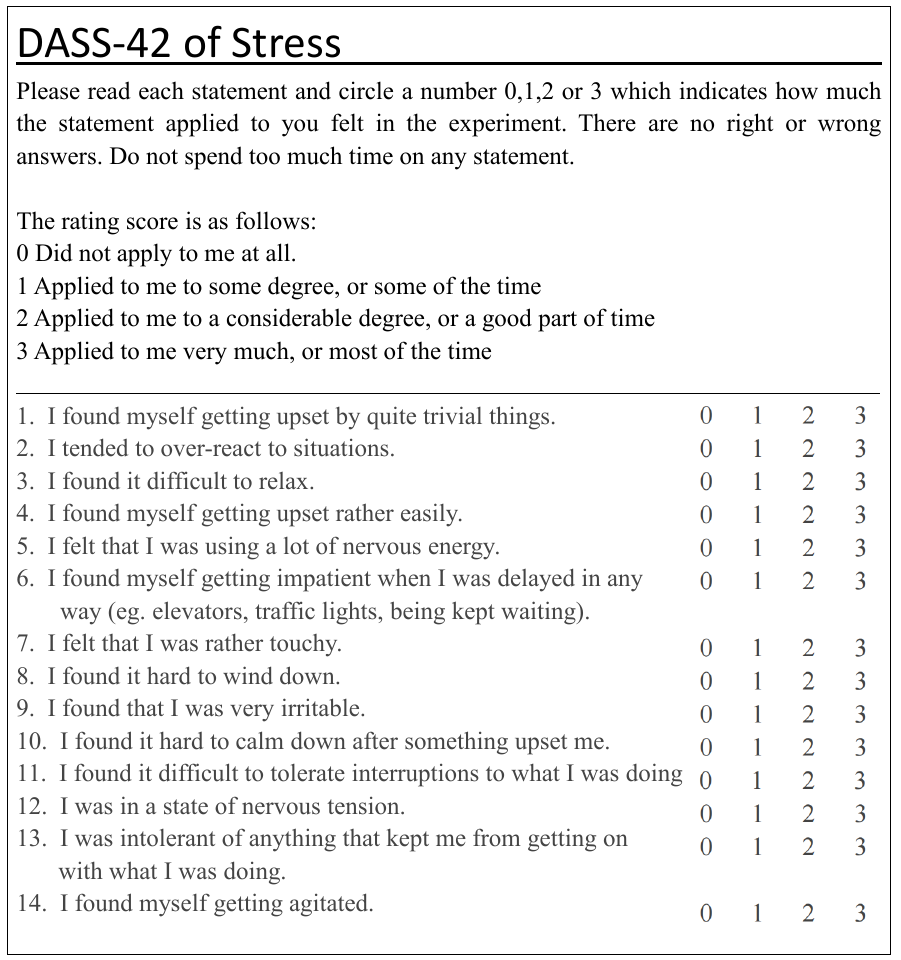}
\caption{Stress subscale of \rev{the} DASS-42}
\label{fig:stress subscale}
\end{figure}
\end{centering}

To demonstrate the effectiveness of the proposed stress-inducing protocol, we use\rev{d} the \textit{Depression Anxiety Stress Scales} with 42-questions (DASS-42) after the target user \rev{completed} the experiment. 
Specifically, \rev{the} DASS-42 consists of three subscales\rev{, namely,} \textit{Stress}, \textit{Anxiety}, and \textit{Depression}, with a total of 42 questions. The three subscales \rev{of the} DASS-42 \rev{examine} the degree \rev{to which} the target \rev{user experiences} stress, anxiety, and depression. In our setting, only the \textit{Stress} subscale in \rev{the} DASS-42 is considered due to the topic \rev{discussed} in this study, where the higher the score given by the target user is, the \rev{greater the} stress state \rev{of} the user. \rev{The contents} of the \textit{Stress} subscale of \rev{the} DASS-42 are shown in Figure \ref{fig:stress subscale}.


\subsection{Experimental Setup}
\subsubsection{Dataset Collection}
\begin{centering}
        \begin{figure}[!t]
            \includegraphics[width=1\linewidth]{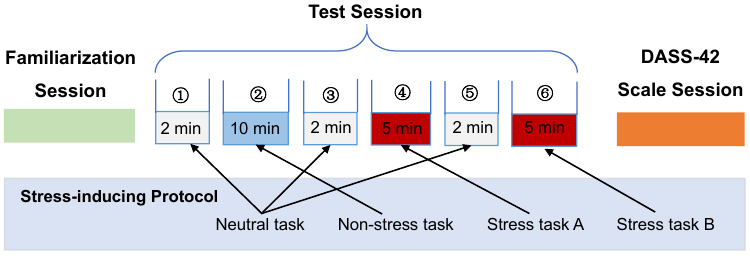}
            \caption{\rev{The} data collection \rev{procedure was} based on \rev{a} stress-inducing protocol (min: minutes)}
            \label{procedure}
        \end{figure}
    \end{centering}

To collect the dataset \rev{that} \rev{contains} multimodal signals captured by a UWB radar, we design\rev{ed} a data collection activity based on the proposed stress-inducing protocol in Section \ref{subsec:proto}. \rev{Table~\ref{Demographics} shows demographics of all target users (i.e., participants) involved in our experiments.}  
All participants report\rev{ed} normal vision and hearing and none of them report\rev{ed} taking any medication or receiving any medication or psychological treatment. 
\rev{Specifically}, the procedure of the data collection activity contains three sessions, which are illustrated in Figure \ref{procedure} and elaborated below. 

\begin{table}[!t]
\begin{center}
\caption{\rev{Demographics of target users involved in experiments}}
\label{Demographics}
\begin{tabular}{|c|c|}
\hline
Demographics & Information  \\ \hline
Gender & 6 females, 11 males         \\ \hline
Age & 17 to 30 years       \\ \hline
Height & 153 to 187 cm         \\ \hline
Weight & 44 to 91 kg        \\ \hline
Occupation & Various industries        \\ \hline
Health background & Healthy         \\ \hline
\end{tabular}
\end{center}
\end{table}


\begin{itemize}
    \item \textit{\textbf{Familiarization Session}}: As shown in Figure \ref{procedure}, in the familiarization session, the purpose and procedure of the data collection activity are explained to each participant. Since each task in the proposed stress-inducing protocol (i.e., neutral task, non-stress task, or stress task) is assigned to the participants in an online test system, the participants are given a short demonstration to understand \rev{how to perform} each task. Note that the order and duration of each task are the same for each participant. \rev{The} participants are informed that a UWB radar is used in the activity to collect their physiological data. At the end of \rev{the} familiarization session, a UWB radar \rev{sensor} is placed on a table in front of each participant, after which a one-minute data acquisition test is conducted to verify \rev{proper operation of} the data collection function of the UWB radar. 

    \item \textit{\textbf{Test Session}}: Figure \ref{procedure} \rev{shows} that the test session involves a sequence of tasks, including \rev{the} neutral task, non-stress task and stress task, which are defined in the stress-inducing protocol. \rev{Specifically}, this session starts with a neutral task that lasts two minutes, which is followed by a non-stress task that lasts 10 minutes. \rev{Afterward}, the neutral task (lasting 2 minutes) and each stress task (lasting 5 minutes) appear\rev{ed} alternately in the session.  
    As \rev{previously} mentioned, each participant \rev{was} supposed to correctly answer as many questions as possible in the time allotted to them during the stress task. 
    Figure \ref{stress task A} and Figure \ref{stress task B} show an example question set for task A and task B in our experiment, \rev{respectively}. 
    The ground-truth labels of the stress state of the target user in stress task A and stress task B are labeled as 1 (indicating under stress), while the ground-truth labels of the stress state of the target user in the non-stress task are labeled as 0 (indicating no stress).

 

    \begin{centering}
        \begin{figure}[h]
        \centering
            \subfloat[\footnotesize An example of question in stress task A]{%
            \includegraphics[width=0.8\linewidth]{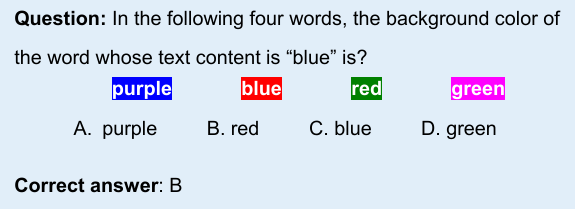}%
              \label{stress task A}%
            }\hfill
            \subfloat[\footnotesize An example of question in stress task B]{%
              \includegraphics[width=0.8\linewidth]{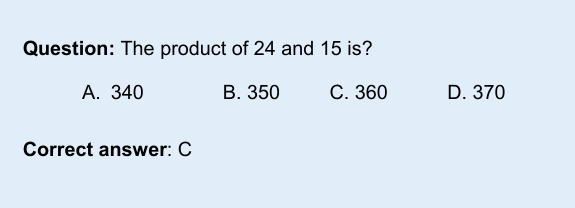}%
              \label{stress task B}%
            }
            \caption{Examples of questions in \rev{the} stress tasks}
            \label{stress task}
        \end{figure}
    \end{centering}
 
    \item \textit{\textbf{DASS-42 Scale Session}}: To verify the effectiveness of the proposed stress-inducing protocol, the DASS-42 scale \rev{was used for} the data collection activity. \rev{The} DASS-42 consists of 42 questions, and each participant is only asked to complete the \textit{Stress} subscale of \rev{the} DASS-42\rev{,} which contains 14 questions with a 4-point Likert scale. 
    The higher the score on the stress subscale of \rev{the} DASS-42 is, the \rev{greater} the stress level \rev{of the target user}. 
\rev{In addition}, we evaluate\rev{d} the effectiveness of the stress-inducing protocol by analyzing the average and standard deviation of the target users' scores on each item of the stress subscale of \rev{the} DASS-42.
\end{itemize}

Figure \ref{experimenter} shows a real-world scene of the data collection activity, where the target user works on a stress task and the UWB radar that is placed in front of the user continuously collects the user's multimodal physiological signal data. To obtain the multimodal dataset collected by the UWB radar, 17 target users are invited to participate into the study. \rev{Specifically}, there are 6 female target users and 11 male target users, ranging in age from 17 to 30 years. Each participator is required to act as the target user and complete the experiment according to the experiment settings, and the experiment is completed once in each of the three scenarios (i.e., classroom, conference room, and bedroom).


\subsubsection{Datasets}
We evaluated the performance of DST using the two public\rev{ly} available datasets most frequently used in stress detection tasks and a new multimodal stress detection dataset that we collect\rev{ed} in the real-world. Table \ref{datasets} shows the description\rev{s} of the three datasets.
The WESAD dataset~\cite{schmidt2018introducing} is the most frequently \rev{employed} dataset in the field of stress detection and can be used to evaluate the performance of \rev{a} model. According to the setting of stress categories, the WESAD dataset includes the two-category WESAD$^2$ dataset and the three-category WESAD$^3$ dataset.
Compared to these two datasets, our dataset provides richer information of modalities using both physiological \rev{signals} and physical signals. Furthermore, unlike the above two datasets collected via wearable sensors, our dataset uses \rev{a} contactless sensor (i.e., UWB radar), which is more friendly to the target user and does not cause possible interference to the target user's stress state.


\begin{table}
\begin{center}
\caption{Description of the three multimodal datasets (N=RespiBAN, E=Empatica E4, and R=UWB radar\rev{)}}
\label{datasets}
\begin{tabular}{|c|c|c|c|c|}
\hline
Name & Target users  & classes & Samples & Sensors  \\ 
\hline
WESAD$^2$& 15 & 2 &  31082   & N, E      \\  \hline
WESAD$^3$& 15 & 3 &  66304   & N, E      \\ \hline
Self-Collected & 17 & 2 &   60992   & R      \\ 
\hline
\end{tabular}
\end{center}
\end{table}

\textbf{WESAD$^2$ dataset}~\cite{schmidt2018introducing}. This dataset contains multimodal physiological data collected from 15 target users which are collected by two human wearable sensors. It can be used for the binary (i.e., stress and non-stress) stress detection task. \rev{Specifically}, a RespiBAN sensor is used to collect the data which has a sampling rate of 700Hz and 8 sensing channels, while another sensor used for collecting data, i.e., Empatica E4, has a sampling rate of 64Hz and 6 sensing channels. In our experiment, we choose a 1-second wide time window to generate a feature matrix based on the data collected by each sensor and deem those data generated by a different sensor are data from different modality. 
We use the data collected from 12 target users for training a human stress detection model and employ the remaining data collected from the other 3 target users for testing the performance of the model in handing stress tracing tasks.

\textbf{WESAD$^3$ dataset}~\cite{schmidt2018introducing}. The WESAD$^3$ dataset follows most of the settings of the WESAD$^2$ dataset. Being different from the WESAD$^2$ dataset, each sample in the WESAD$^3$ dataset is labeled with a more detailed stress state (i.e., baseline, stress, and amusement), rather than only two categories of stress states (i.e., stress and non-stress) used by the WESAD$^2$ dataset. Therefore, the WESAD$^3$ dataset can be used to evaluate the performance of different stress detection models in distinguishing more stress categories.

\begin{figure}[h]%
    \centering	\includegraphics[width=.79\columnwidth]{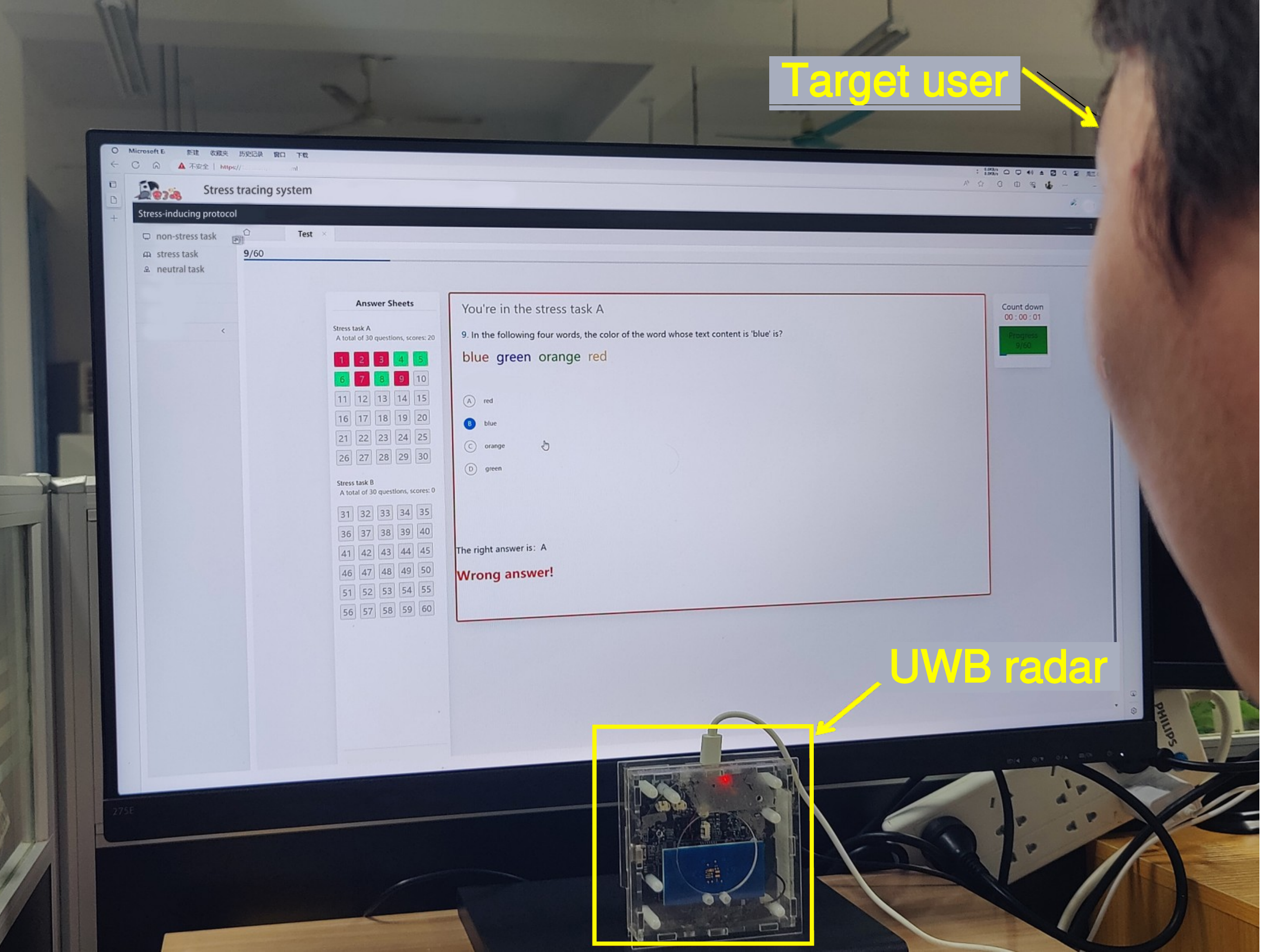}
    \caption{A real-world scene of the data collection activity}\label{experimenter}
\end{figure}

\textbf{Self-collected dataset.}
It is a self-collected real-world dataset which is derived following the stress-inducing protocol proposed in Section \ref{subsec:proto}. \rev{Specifically}, this dataset contains multimodal physiological signals of 17 target users (11 males and 6 females) collected by a UWB radar deployed in three different indoor environments, including classroom, conference room, and bedroom. A typical scene for data collection is shown in Figure \ref{experimenter}. 
We choose a 5-second wide sliding time window to collect RF data and each sample in the dataset is labeled with a binary stress state, where 1 indicates under stress and 0 represents non-stress. \rev{Specifically}, the self-collected data contains two modalities of data, namely heart rate (HR) and respiratory rate (RR). We use the data collected from 14 target users for training a human stress detection model and employ the remaining data collected from the other 3 target users for testing the performance of different models in handing a stress tracing task.

\subsubsection{Baselines}
The effectiveness of the proposed DST method is compared with \rev{that of ten}  baselines\rev{,} all of which are vital physiological\rev{-}based methods for human stress detection. We list all comparison baselines and their descriptions below. 
\begin{itemize}[itemsep=0pt] 
    \item \textbf{SD\_AE}~\cite{rodriguez2020towards}: it proposes to use the physiological features to predict human stress in an academic environment. \rev{Specifically}, SD\_AE employs the $k$-nearest neighbor algorithm to select the most robust physiological feature and uses it for predicting human stress states.
    \item \textbf{RTEEG}~\cite{alshorman2022frontal}: it utilizes the features extracted from a single physiological signal via the Fast Fourier Transform as the input, and employs the support vector machine for human stress detection.
    \item \textbf{PSNS}~\cite{masood2019modeling}: it concatenates multiple physiological signals from two modalities through the mathematical pre-processing, and then uses multiple convolutional layers and recurrent neural networks (RNN) with LSTM layers to detect stress of the target user.
    \item \textbf{DeepCNN}~\cite{li2020stress}: it extracts features from multiple physiological signals using a 1D-CNN and treats the concatenation of all feature vectors as its input. Then, multiple fully connected layers are utilized in DeepCNN to achieve the prediction of human stress states.
    \item \textbf{Mostress}~\cite{de2022mostress}: it uses a time-series-based RNN structure to achieve the stress detection. Note that Mostress makes use of the mathematical computations on the physiological data collected by wearable devices to improve the performance of stress detection. 
    \item \textbf{IEBDNN}~\cite{ghosh2022classification}: it simply connects various physiological signals to get the fusion feature in the pre-processing stage, and then encodes those fusion features into GAF images which are used to train a CNN model for human stress detection.
    \item \textbf{WiStress}~\cite{ha2021wistress}: it proposes \rev{using} the uni-modal physiological signal of a target user\rev{,} which is collected by an mmWave radar\rev{,} to achieve the goal of human stress detection. WiStress sends mmWave signals to a target user, analyzes the signals reflected by the user to estimate his \rev{or} her heartbeat, and infers the user's stress state based on the heartbeat information.
    \item \textbf{mmStress}~\cite{liang2023mmstress}: it is the latest method for stress detection based on wireless sensing. mmStress predicts human stress by recognizing and analyzing displacement activities under stress (e.g., walking around and scratching) \rev{at} \rev{macro- and micro-temporal} scales via a customized neural network.
    \item \rev{\textbf{WiStress+SEM}: it is a variant of WiStress where the proposed SEM in DST is used as its signal extraction module.}
    \item \rev{\textbf{mmStress+SEM}: it is a variant of mmStress where the proposed SEM is used as its signal extraction module.}
\end{itemize}

\subsubsection{Evaluation Metrics}
As to the domain of human stress detection, the most commonly used evaluation metrics are accuracy (ACC), F1 score, and area under the curve (AUC). Here, ACC is defined as the proportion of samples with correct predictions over all samples, F1 score is a statistical metric used to measure the accuracy of a binary classification model, and AUC is defined as the area enclosed by the ROC curve and the lower coordinate axis. 
Generally speaking, the value 0.5 for AUC or ACC indicates that the classification (or detection) performance is the same as the random guessing, and the value 1 for AUC or ACC signals that the detection model achieves perfect performance. Note that the AUC generally takes a value from the range of $[0,1]$ and the closer the AUC value is to 1, the better the stress detection performance is. F1 score considers both the precision and recall of the classification model, and it offers a balanced assessment towards a detection model's performance in classifying positive samples and negative samples. For each method, we repeat the training process five times and report the average ACC, F1 score, and AUC in the following experiments.

\begin{table*}[t]
\begin{center}
\caption{Implementation details of multimodal fusion module (MFM) in the proposed DST method (Parameters of cross connections are shown in \textbf{bold}, $\searrow$ denotes 1D$\to$3D cross connection, and $\swarrow$ denotes 3D$\to$1D cross connection).}\label{parameter}
\begin{tabular}{|c|c|c|c|}
\hline
 HR signal stream &  RR signal stream & RF embedding stream  &  Output size\\ \hline
 
 Fully-connected 32D & Fully-connected 32D & Conv. 2D [3 × 3, 8]& ([150 × 96, 8], 32) \\ \cline{4-4}
 & &  Max-pool. [2 × 2] & ([75 × 48, 8], 32) \\ \hline
 
 \textbf{Fully-connected 192D} & \textbf{Fully-connected 192D} & \textbf{Conv. 2D [1 × 1, 8]} & \multirow{2}{*}{\textbf{([75 × 48, 16], 96)}} \\ 
 \textbf{Deconv. 2D [16 × 12, 8]} $\searrow$ & \textbf{Deconv. 2D [16 × 12, 8]}$\searrow$ & $\swarrow $ \textbf{Fully-connected 32D} &  \\ \hline
 
 Fully-connected 64D & Fully-connected 64D & Conv. 2D [3 × 3, 24] & ([75 × 48, 24], 64) \\ \cline{4-4}
 & & Max-pool. [2 × 2] & ([37 × 24, 24], 64) \\ \hline

  \textbf{Fully-connected 40D} & \textbf{Fully-connected 40D} & \textbf{Conv. 2D [1 × 1, 24]} & \multirow{2}{*}{\textbf{([37 × 24, 72], 192)}} \\ 
 \textbf{Deconv. 2D [8 × 5, 24] $\searrow$} & $\swarrow $ \textbf{Deconv. 2D [8 × 5, 24]} & $\swarrow $ \textbf{Fully-connected 64D}&  \\ \hline

 Fully-connected 192D &  &  & (192, 64) \\ \hline
\end{tabular}
\end{center}
\end{table*}

\subsubsection{Radar Selection}
\rev{
Advantages and disadvantages of two representative types of radars (i.e., FMCW radar and UWB radar) for vital signs detection tasks are usually compared by researchers~\cite{wang2020experimental}. Resorting to the power of wide-band, both of the FMCW radar and UWB radar achieve good detection performance due to their high time resolution. However, the UWB radar slightly outperforms FMCW radar since it: i) has fewer harmonics and a higher signal-to-noise ratio (SNR)~\cite{wang2020experimental}; ii) is immune to multipath distortions due to its characteristics of short pulse signals~\cite{luo2012indoor} while the FMCW radar is susceptible to jamming in the presence of multipath propagation~\cite{lu2016fmcw}. 
Another popular type of radar is the millimeter-Wave radar (mmWave). We do not choose the mmWave radar in our study to collect physiological signals of target users for the following two reasons.
First, the mmWave radar is sensitive to the influence of air humidity conditions. In specific, under the condition of a high atmospheric humidity, the transmitted signal of mmWave radar is easy to be scattered and absorbed, which reduces the detection and ranging ability of the mmWave radar~\cite{dai2018mm}. In contrast, the UWB radar maintains a good performance under most conditions of air humidity. 
Second, the mmWave radar has a lower bandwidth and SNR compared to the UWB radar, which makes it have a relatively weak anti-jamming ability 
than the UWB radar~\cite{wang2020experimental}.
It is also noteworthy that due to the wide frequency spectrum distribution of UWB signals that occupies a large frequency range, the UWB radar has a superior immunity to on-air interference from other wireless devices operating in the same frequency band~\cite{UWB23}\cite{chen2022harris}. In other words, the UWB radar can coexist with other communication systems, e.g., Wi-Fi and Bluetooth, without causing significant interference, making it a reliable and robust choice for even those applications that require uninterrupted operations. It is those above-mentioned advantages of the UWB radar that make us select the UWB radar as the basic support for implementing our proposed stress tracing method. 
}

\subsubsection{Scenario Settings}
\rev{To ensure the UWB radar can effectively detect the stress states of each target user, we require that: i) the position of the target user is within the Field of View (FoV) of the UWB radar so that the radar can receive the reflected signal from the target user, ii) the distance between the radar and the user (i.e., monitoring distance) lies within a reasonable sensing range (generally 0.3 m to 3 m~\cite{zheng2021more}) to alleviate the multipath effect of the UWB radar, and iii) only the body movements that away from the user's chest (e.g., typewriting, limb position drift, drinking water, and yawning) rather than those movements drastically changing the user's posture (e.g., from sitting to lying) are allowed to ensure the monitoring of respiration and heart rate of the user is not affected.}

\subsubsection{Other Settings}
For the signal extraction module (SEM) in the proposed DST method, the dimensionality of each of the two fully connected layers in the self-attention layer of SEM is set to 64 to facilitate the signal extraction. Note that the \textit{ReLU} activation function is applied to all convolutional and fully-connected layers in DST. The multimodal fusion module (MFM) is the most complicated module in DST, since different modalities of signals and their interactions are processed and learned in the module. 
Table \ref{parameter} summarizes the implementation details of MFM, where the description of residual connections in MFM is omitted for they can be inferred from the shape of the output.
To combat overfitting, a couple of regularization techniques are employed in MFM. \rev{Specifically}, the batch normalization technique is applied after the input layer (or the operation that fuses ${{\boldsymbol{\tilde{v}}}_{hr}}, {{\boldsymbol{\tilde{v}}}_{rr}}$, and ${{\boldsymbol{\tilde{v}}}_{rf}}$) in MFM, which normalizes the input to each layer, reduces internal covariate shift, and improves model generalization.
Besides, the dropout technique is applied after every max-pooling layer in MFM with a dropout rate of $0.25$. Notice that a larger dropout rate of $0.5$ is applied after the first fully-connected layer in MFM. This is because the fully-connected layer has more parameters than each of the max-pooing layers and a larger dropout rate offers more help to counteract overfitting for the fully-connected layer.
For the stress tracing module (STM) in DST, the size of the hidden layer of the LSTM is set to 64, and the LSTM layer is followed by a fully-connected layer with a 2-way softmax activation function for stress tracing. It is worth noting that the dropout rate in the LSTM layer is set to $0.25$. 

To evaluate each comparison model, 80\% of the collected dataset are used as the training set and the remaining 20\% are treated as the test set. During the training stage of every model, we use Adam with a batch size of 40, and run the experiments with 50 epochs, while the learning rate is set to 0.001 for all comparison methods. 
We use a commercial off-the-shelf UWB radar from AIWise~\cite{wirush2023} to collect experimental data. The radar operates at 7.3 GHz with a bandwidth of 1.4 GHz and 150 frames per second (FPS); it has a pair of Tx-Rx antennas with a field of view of 65° angle. 
In addition, all comparison methods are implemented by PyTorch 2.1, and trained using the same hyper-parameters on a Linux server with an Intel Core i9-11900 CPU (8 cores), a NVIDIA RTX A5000 GPU, and 128 GB DDR4 RAM.
To ensure the fairness of comparison, all baselines compared in the following experiments are tuned to their best performance.

\subsection{Experimental Analysis}
\subsubsection{Hyper-parameter sensitivity}
In this section, we provide a sensitivity analysis for two hyper-parameters which are used by the proposed DST method.
\begin{itemize}[itemsep=0pt] 
\item \textbf{Impact of the number of convolutional layers in the CNN layer of SEM.} Figure \ref{parameter_sensitivity:CNN result} displays the impact of different number of convolutional layers in the CNN layer of SEM (denoted as $N_{CNN}$) on the performance of DST on the self-collected dataset, when the number of heads set for the self-attention mechanism in SEM of DST (denoted as $N_{head}$) is fixed to 4. As shown in the figure, with the increase of $N_{CNN}$, the value of ACC, F1 score, or AUC displays a gradual downward trend. 
This may be because each convolutional layer transforms the information, and as the number of convolutional layers increases, the time-domain and frequency-domain information in the input data will gradually be lost, making DST be unable to effectively capture important features in the input data.
Thus, $N_{CNN}$ is set to 1 in DST on self-collected dataset in the following experiments, to ensure that DST achieves best performance. On WESAD$^2$ and WESAD$^3$ datasets, CNNs are used to encode features from the inputted multimodal data. Figures \ref{parameter_sensitivity:CNN result-1} and \ref{parameter_sensitivity:CNN result-2} show the impact of different $N_{CNN}$ on the two datasets separately, when $N_{head}$ is fixed to 4. As shown by the two figures, DST consistently gains its best performance when $N_{CNN}$ equals to 3, due to the over-fitting phenomena.
Therefore, $N_{CNN}$ is set to 3 on the WESAD$^2$ and WESAD$^3$ datasets in the following experiments. 

\begin{figure*}[h]
    \centering
    \subfloat[\footnotesize Impact of $N_{CNN}$ on self-collected dataset]{
    \includegraphics[width=0.31\linewidth]{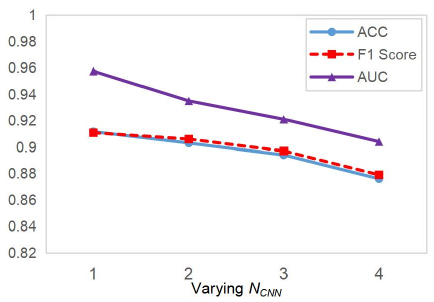}
    \label{parameter_sensitivity:CNN result}
    }
    \hfill
    \subfloat[\footnotesize Impact of $N_{CNN}$ on WESAD$^2$ dataset]{
        \includegraphics[width=0.31\linewidth]{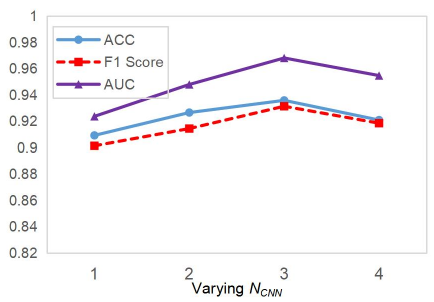}
        \label{parameter_sensitivity:CNN result-1}
    }
    \hfill
    \subfloat[\footnotesize Impact of $N_{CNN}$ on WESAD$^3$ dataset]{
    \includegraphics[width=0.31\linewidth]{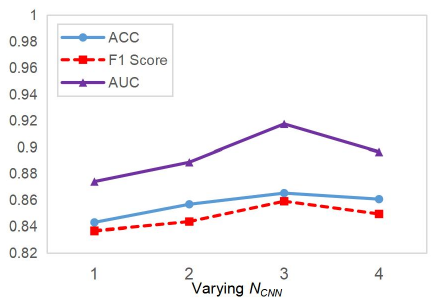}
    \label{parameter_sensitivity:CNN result-2}
    }
    \caption{Sensitivity analysis of the number of convolutional layers ($N_{CNN}$) on three datasets}
    \label{parameter_sensitivity:CNN}
\end{figure*}

\begin{figure*}[h]
    \centering
    \subfloat[\footnotesize Impact of $N_{head}$ on self-collected dataset]{
        \includegraphics[width=0.31\linewidth]{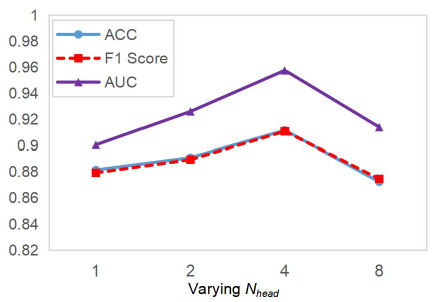}
        \label{parameter_sensitivity:head result}
    }
    \hfill
    \subfloat[\footnotesize Impact of $N_{head}$ on WESAD$^2$ dataset]{
        \includegraphics[width=0.31\linewidth]{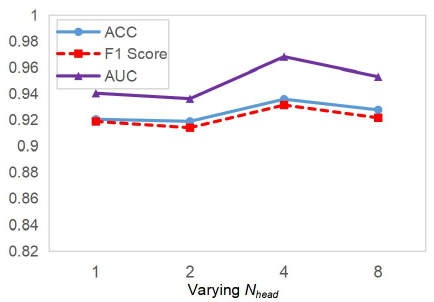}
        \label{parameter_sensitivity:head result-1}
    }
    \hfill
    \subfloat[\footnotesize Impact of $N_{head}$ on WESAD$^3$ dataset]{
        \includegraphics[width=0.31\linewidth]{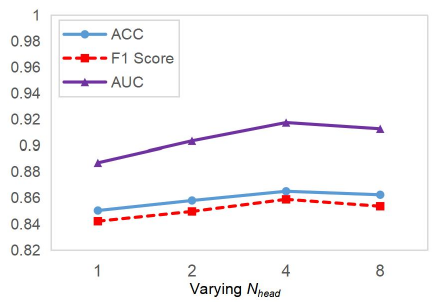}
        \label{parameter_sensitivity:head result-2}
    }
    \caption{Sensitivity analysis of the number of heads in self-attention ($N_{head}$) on three datasets}
    \label{parameter_sensitivity:head}
\end{figure*}

\item \textbf{Impact of the number of heads in SEM. } Figures \ref{parameter_sensitivity:head} investigates the impact of different amount of head (i.e., $N_{head}$) set for the self-attention mechanism in SEM on the prediction performance of DST on three datasets, while the number of convolutional layers $N_{CNN}$ is fixed to $1$. As demonstrated by these three figures, when $N_{head}$ equals to 4, DST achieves its best performance in terms of ACC, F1 score, and AUC, while introducing more heads beyond 4 degrades the performance of DST. Therefore, $N_{head}$ is set to 4 in the following experiment in DST on all datasets in the following experiments, to ensure DST can derive its best prediction results. 
\end{itemize}

\subsubsection{Comparison with Baselines}
Table \ref{comparisons} displays the experimental results of comparing our proposal (i.e., DST) with baselines. Overall, the average improvement\rev{s} in terms of \rev{the} ACC, F1, and AUC of DST compared with \rev{those of} best baselines on three datasets \rev{are 6.31\%, 7.24\%, and 6.69\%}, respectively. 
\rev{DST significantly outperforms all baselines in terms of ACC, F1, and AUC, which demonstrates} the superiority of the proposed DST method in extracting and utilizing the multimodal physiological signals of target users when tracing their stress states. \rev{Note that two variants of WiStress and mmStress, i.e., WiStress+SEM and mmStress+SEM, are only evaluated on the self-collected dataset, since the embedded SEM in these two variants is designed for extracting HR or RR signals from the raw RF data collected by a UWB radar while the HR or RR data have already been provided in the WESAD$^2$ or WESAD$^3$ dataset.}
Next, we \rev{provide a} detailed analysis of \rev{the} results shown in Table \ref{comparisons}. \rev{Specifically}, by observing the table, we derive the following findings. 

\begin{table*}
\begin{center}
\caption{A comparison between DST and baselines. 
Data mark\rev{ed} with * denote the next best results. The last column gives the improvement rate gained by DST\rev{s} compared with the next best result\rev{s}.}
\label{comparisons}
\begin{tabular}{|c|c|c|c|c|c|c|c|c|c|}
\hline
\multirow{3}{*}{Model}  & \multicolumn{3}{c|}{WESAD$^2$ Dataset} &
\multicolumn{3}{c|}{WESAD$^3$ Dataset} & \multicolumn{3}{c|}{Self-collected Dataset} \\   \cline{2-4} \cline{5-7} \cline{8-10}  

& Accuracy & AUC & F1-score & Accuracy & AUC  & F1-score & Accuracy & AUC & F1-score \\
& (\%) & (\%) & (\%) & (\%) & (\%)  & (\%) & (\%) & (\%) & (\%) \\ 
\hline

SD\_AE~\cite{rodriguez2020towards}&  72.87   &  76.43   &  72.12  &   64.35  &     67.04 & 63.56   &  66.77   &  71.04   &  64.37    \\ \hline
RTEEG~\cite{alshorman2022frontal}&  70.63   &  72.15   &  70.54  &  63.69   &  64.40   & 63.17   &  69.29   &  74.25   &  68.81    \\ \hline
PSNS~\cite{masood2019modeling}& 80.71  &	84.18	 &  80.36	 &  73.17  & 	77.03	 &  72.46	& 72.14	 &  76.47	 &  70.01      \\ \hline
DeepCNN~\cite{li2020stress}& 86.35 & 89.07 & 85.44 & 75.32	& 79.84	& 74.95	& 73.15 & 78.30 & 73.17      \\ \hline
Mostress~\cite{de2022mostress}& 81.54 & 84.91 & 80.23 & 74.58 & 78.16 & 72.91 & 73.35 & 79.78 & 74.69  \\ \hline
IEBDNN~\cite{ghosh2022classification}& 84.32 &	88.68 & 84.15 & 78.43 & 81.57 & 77.64 & 75.59 & 78.85 & 73.42     \\ \hline
WiStress~\cite{ha2021wistress}& 83.46 & 86.72 & 83.21 & 77.62 & 81.73 & 77.08 & 84.83 & 88.36 & 83.24     \\ \hline
mmStress~\cite{liang2023mmstress}& 86.59*	& 89.57* & 86.37* & 79.19*	& 83.28* & 78.41* & 80.79 & 85.02 & 79.53      \\ \hline

\rev{WiStress+SEM}& \rev{---}	& \rev{---} & \rev{---} & \rev{---} & \rev{---} & \rev{---} & \rev{85.61} & \rev{88.75} & \rev{84.41}    \\ \hline
\rev{mmStress+SEM}& \rev{---} & \rev{---} & \rev{---} & \rev{---} & \rev{---} & \rev{---} & \rev{86.53*}  & \rev{89.72*}  & \rev{85.26*}   \\ \hline

\textbf{DST (Ours)} & \textbf{93.58} & \textbf{96.81} & \textbf{93.13} & \textbf{86.51} & \textbf{91.75} & \textbf{85.89} & \textbf{91.16} & \textbf{95.75} & \textbf{91.08}  \\  \hline
Improvement &  6.99   &  7.24   &  6.76  &  7.32   &  8.47   &  7.48  &      \rev{4.63} &    \rev{6.03}  & \rev{5.82} \\
\hline
\end{tabular}
\end{center}
\end{table*}

\begin{itemize}
    \item \textbf{Finding 1}: \textit{Deep learning based methods obtain the most accurate stress detection result.} As shown in Table \ref{comparisons}, methods that use the traditional machine learning strategies (i.e., SD\_AE and RTEEG) are beaten by those deep learning based methods (i.e., PSNS, DeepCNN, Mostress, IEBDNN, WiStress, mmStress, \rev{WiStress+SEM, mmStress+SEM,} and DST).
    This is because deep learning based methods own significant advantages in learning high-quality representations of multimodal physiological signals and they thus can effectively capture the nonlinear relationships between multimodal signals and a target user's stress states. 
    
    \item \textbf{Finding 2}: \textit{Raw physiological signal data need to be processed before they can be used to tracing human stress.}
    As displayed in Table \ref{comparisons}, on all datasets, methods that use the processed physiological signal data (i.e., RTEEG, DeepCNN, Mostress, IEBDNN, WiStress, mmStress, \rev{WiStress+SEM}, \rev{mmStress+SEM}, and DST) are consistently superior to those methods that directly use the raw physiological signal data (i.e., SD\_AE and PSNS) in detecting human stress. This is because the collected raw physiological signal data generally contain noises from background or unconscious body movements, which brings negative impact on the precision stress detection results. This finding verifies the importance of filtering out the noises in the raw physiological signal data when designing a stress method.
    First, in DST, the background noise is removed from the raw RF data by mean difference of the signal technique. Second, FFT is applied to convert the time domain signal of the RF data where background noise has been filtered out into the frequency domain signal. Furthermore, the signal extraction module (SEM) of DST can extract the reliable HR and RR signals from the raw RF data collected by a UWB radar addressing the two challenges facing by signal extraction, i.e., non-linearity of signals and unintentional body movements of the target user.
    
    \item \textbf{Finding 3}: \textit{Optimizing the signal extraction strategy is vital for ensuring the performance of stress detection.} 
    When extracting representations for physiological signals of different modalities, the characteristics of different modalities should be considered. However, many baselines simply apply the same mathematical calculation method (i.e., Mostress and IEBDNN) or the same convolutional network structure (i.e., DeepCNN and PSNS) to learn the representations for physiological signals of different modalities, leading to weak generalization ability, low interpretability, and unsatisfactory stress detection performance.
    Compared with those baselines, in DST an \rev{effective signal extraction module (SEM) is applied. Specifically, in SEM,} for the two modalities HR and RR, both of a CNN structure and a multi-head attention mechanism are used to extract their representations which reflect the information of time domain, frequency domain, and temporal correlation within each physiological signal stream.
    Moreover, to derive a high-quality representation of the third modality in DST (i.e., RF embedding), DST proposes to combine the raw RF data's real part feature which reflects the signal strength information and the imaginary part feature that contains the signal phase information. 
    As shown in Table \ref{comparisons}, on three datasets, the values of ACC, AUC, and F1 score of DST gain averagely 12.95\%, 13.37\%, and 13.16\% increase separately, compared with Mostress, IEBDNN, DeepCNN, and PSNS which only use less-satisfactory strategies to extract signals. \rev{Notice that he effectiveness of the proposed signal extraction strategy (i.e., SEM) in DST can be also verified by comparing WiStress and mmStress with their variants, i.e., WiStress+SEM and mmStress+SEM. In particular, with the application of SEM, these two methods averagely gain performance increase of $0.78\%$ and $5.74\%$ respectively on ACC on the self-collected dataset.}

    \item \textbf{Finding 4}: \textit{Considering temporal correlation of physiological signals can significantly improve the detection performance}. As shown in Table \ref{comparisons}, DST, mmStress, \rev{and mmStress+SEM} which can be deemed as the stress tracing methods rather than the stress detection methods perform consistently better than other baselines on all datasets. This is because these \rev{three} methods consider the temporal correlation of signals derived by applying the self-attention mechanism over each input signal stream. It is the consideration of temporal correlation within a signal stream that makes DST, mmStress, \rev{and mmStress+SEM} better trace human stress and therefore derive more accurate stress state prediction results.

    \item \textbf{Finding 5}: \textit{Fusing multimodal physiological signals are very important to improve stress detection results.} Fusing multimodal signals aims to learn a stronger representation than via a single modality. 
    In Table \ref{comparisons}, on all datasets, PSNS that does not utilize the information of multimodal fusion performs worse than other deep learning based baselines that make use of the multimodal information (i.e., DeepCNN, Mostress, IEBDNN, WiStress, mmStress\rev{, WiStress+SEM, and mmStress+SEM}). Another observation from Table \ref{comparisons} is that the proposed DST method performs averagely 8.45\% and 8.23\% better in accuracy compared with WiStress and mmStress separately on all datasets. Notice that WiStress and mmStress are the state-of-the-art non-contact solutions for human stress detection. The superiority of DST comes from the consideration of information exchange among different modalities of sensing signals, while WiStress and mmStress simply concatenate signals of different modalities when training their models. 

        %
        
    \item \textbf{Finding 6}: \textit{DST dramatically \rev{outperformed} all \rev{the} baselines}. 
    DST \rev{adequately} handles the two challenges of extracting physiological signals by considering the information \rev{in the} time domain and frequency domain of RF signals, and effectively learns the temporal correlation within a single signal via a multi-head self-attention mechanism. \rev{Moreover,} cross connection\rev{s} with residual connection\rev{s} \rev{are} used to directly exchange information in the physiological signals of different modalities to better \rev{utilize} the correlation and complementarity between different modalities, which can improve the ability of DST\rev{s} \rev{to fuse} multimodal signals. 
    Another observation from Table \ref{comparisons} is that DST performs better on \rev{the} WESAD$^2$ dataset than \rev{on the} other two datasets. \rev{On} one hand the WESAD$^2$ dataset is collected by wearable sensors and thus contains relatively less noise than the self-collected dataset\rev{,} which is collected using non-contact UWB radar. On the other hand, the stress state labels defined on \rev{the} WESAD$^2$ dataset only have two categories (i.e., under stress and non-stress), while the stress labels defined on \rev{the} WESAD$^3$ dataset have three categories (i.e., baseline, stress, and amusement), which makes the stress tracing task on WESAD$^3$ dataset more complicated than that on \rev{the} WESAD$^2$ dataset.

    \item \textbf{Finding 7}: \textit{DST has a good generalization in handling different sensing data for stress detection.} The significant performance improvement of DST gained on the two publicity datasets (i.e., WESAD$^2$ and WESAD$^3$) compared with baselines also show that DST has a good generalization in handling different multimodal sensing datasets for human stress detection. Notice that both of the WESAD$^2$ and WESAD$^3$ datasets are collected through human wearable sensors rather than non-contact radars. The superiority of DST gained on this two datasets, once again, confirms the effectiveness of DST in fusing and utilizing different multimodal sensing data. 


\end{itemize}

\begin{table*}[!t]
\begin{center}
\caption{Results of Ablation Study on self-collected dataset (Best values are in bold)}
\label{variants-result}
\begin{tabular}{|c|c|c|c|c|c|c|c|c|}
\hline
Metrics & DST-SEM & DST-HR & DST-RR & DST-X-R & DST-RF-X-R & DST-STM  & DST \\ \hline
ACC & 0.8137  & 0.8836  & 0.8831 &  0.8247   & 0.8336  & 0.8710  & \textbf{0.9118}    \\ \hline
F1 score& 0.8019  & 0.8823  & 0.8847 &  0.8239   & 0.8381  & 0.8763  & \textbf{0.9108}    \\ \hline
AUC & 0.8055  & 0.8975  & 0.9020 &   0.8385    & 0.8423  & 0.8824 & \textbf{0.9575}    \\ \hline
\end{tabular}
\end{center}
\end{table*}


\subsubsection{Ablation Study}
To analyze the effectiveness of the proposed modules in DST in coping with stress tracing tasks, we perform an ablation study by comparing DST with its six variants on the self-collected dataset. For a fair comparison, all of the comparison methods are tuned to achieve their best performance. Detailed descriptions of every DST variants are provided below. 
\begin{itemize}
    \item \textbf{DST-SEM}: it is a variant of DST that removes the signal extraction module (SEM) and uses a mathematical calculation algorithm~\cite{liang2018improved} to extract HR and RR signals.
    \item \textbf{DST-HR}: it is a variant of DST that eliminates the HR signal stream in Multimodal Fusion Module (MFM).
    \item \textbf{DST-RR}: it is a variant of DST which removes the RR signal stream in MFM.
    \item \textbf{DST-X-R}: it is a variant of DST that cuts off both of the cross connection (i.e., $X_{conn}$) and residual connection (i.e., $R_{conn}$) in MFM.
    \item \textbf{DST-RF-X-R}: it is a variant of DST that deletes the RF embedding input of MFM, and removes $X_{conn}$ and $R_{conn}$ from MFM.
    \item \textbf{DST-STM}: it is a variant of DST where the Stress Tracing Module (STM) is eliminated and instead a softmax operation is applied by the variant for predicting the human stress states.
\end{itemize}

Table \ref{variants-result} shows the experimental results of the ablation study. First, we notice that DST outperforms every variant of it, which indicates the effectiveness of each modules or inputs designed for DST.
Second, the performance of DST-SEM has shown a decrease of 9.8\% in ACC, 10.9\% in F1 score, and 15.2\% in AUC compared with DST. This shows the superiority of the proposed signal extraction module (SEM) in DST towards the mathematical calculation algorithm with respect to extract HR and RR signals from the raw RF data collected by the UWB radar. 
\rev{Specifically}, without SEM, it is very difficult for DST-SEM to solve the challenges of signal interference and unintentional body movements when extracting HR and RR signals. 
Third, since the three variants of DST, namely DST-HR, DST-RR, DST-X-R, and DST-RF-X-R, are derived by revising the implementation of the multimodal fusion module (MFM) in DST, the deteriorating performance of the three variants compared with DST demonstrates the importance of MFM in effectively fusing multimodal physiological signals.
In particular, DST-HR and DST-RR perform 2.82\% and 2.85\% worse in ACC than DST separately, which indicates the vital function of HR or RR physiological signal in tracing the stress state change of a target user. 
Meanwhile, we can conclude that the proposed cross connection and residual connection in MFM are very important to ensure the adequate information exchange between a pair of signal streams of different modalities, since without using these two types of connections the two related variants of DST, i.e., DST-X-R and DST-RF-X-R, can only generate less satisfactory prediction results. It is the sufficient information exchange among signals of different modalities that makes DST generate high-quality fused representations of multimodal signal data compared with its variants.  
Notice that DST-X-R retains the RF embedding but achieves worse performance than DST-RF-X-R. \rev{Although} RF embedding contains richer information than HR or RR signals, it also owns lots of redundant interference information. In such case, when information exchange is not performed in MFM (i.e., $X_{conn}$ and $R_{conn}$ are removed from MFM), introducing the RF embedding stream into MFM (w.r.t. DST-X-R) can only bring negative impact on predicting human stress states. 
Finally, we can observe from Table \ref{variants-result} that 
DST-STM performs averagely 5.01\% worse than DST in terms of all evaluation metrics, which indicates that using the LSTM structure to trace human stress states is more effective than using the simple softmax operation. This is because the LSTM structure is very effective to process time series data and maintain temporal integrity.


\subsubsection{\rev{Impact of Different Monitoring Distances}}
\label{sub:distance}
\begin{figure*}[!htb]
    \centering
    \subfloat[\scriptsize \rev{Varying monitoring distance}]{
    \includegraphics[width=0.31\linewidth]{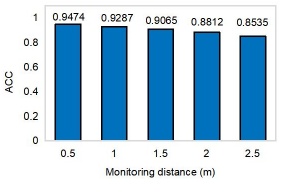}
    \label{distances}
    }
    \hfill
    \subfloat[\scriptsize \rev{Varying radar placement angle}]{
        \includegraphics[width=0.31\linewidth]{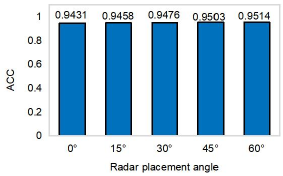}
        \label{placements}
    }
    \hfill
    \subfloat[\scriptsize \rev{\rev{Varying environmental condition}}]{
    \includegraphics[width=0.31\linewidth]{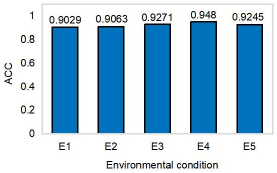}
    \label{environments}
    }
    \caption{\rev{Performance of DST under different settings}}
    \label{conditions}
\end{figure*}

\rev{The monitoring distance, i.e., the distance between radar and monitored target user, is a major factor that affects the performance of radar-based applications. To evaluate the impact of different monitoring distances to our proposal, we have set up five scenarios, where the monitoring distance is set to 0.5 m, 1 m, 1.5 m, 2 m, and 2.5 m, separately. In every scenario, a target user is asked to complement three round of experiments according to the stress induction protocol elaborated before, and physiological signal data of the target user in each experiment are collected. Then, the data collected in the first two experiments are used as the training dataset, while the data gathered in the third experiment are treated as the test dataset. Experimental results of the impact of different monitoring distances on the performance of DST are reported in Figure \ref{distances}. As shown in the figure, with the increase of the monitoring distance, the ACC (accuracy) of DST gradually declines. In particular, the ACC of DST decreases by 9.91\% when distance varies from 0.5 m to 2.5 m. 
This is because with the increase of monitoring distance: i) the difficulty of the UWB radar to distinguish chest vibrations from other body movements raises; ii) and there is a degradation w.r.t. SNR of the RF signals. 
In our default experimental settings, the UWB radar is put under the computer display screen (see Figure~\ref{experimenter}), and thus the monitoring distance generally varies between 0.5 m and 0.6 m, which ensures DST can get its best performance for stress detection tasks. Notice that to ensure the effectiveness of DST in stress tracing (i.e., the stress detection accuracy being larger than 0.9), the maximum monitoring distance is suggested to be 1.5 m.}


\subsubsection{\rev{Impact of Different Radar Placement Angles}}
\label{sub:angle}
\begin{figure}[h]
    \centering
    \includegraphics[width=0.7\linewidth]{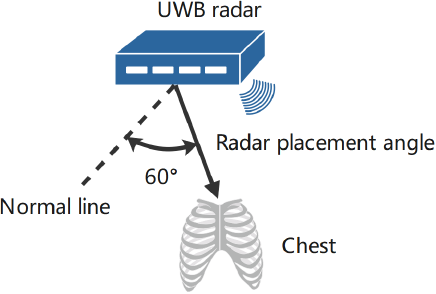}
    \caption{\rev{Illustration of the radar placement angle}}
    \label{angle}
\end{figure}

\rev{In this section, impact of different radar placement angles (angles for short) on the performance of DST is verified. The radar placement angle demonstrated in Figure \ref{angle} is the angle between the radar's normal line and its line pointing to the chest of a target user, and it characterizes a situation of sensor placement (or user face orientation). Specifically, the angle is separately set to 0°, 15°, 30°, 45°, 60° in our experiments reflecting five situations of sensor placement, and the target user is within the radar field of view (FoV) in all situations. In each situation, the target user is required to complete three round of experiments based on our stress induction protocol, and the RF data gathered in the previous two experiments are used to train DST while the data from the last experiment are applied to test DST. Experimental results of varying angles are displayed in Figure \ref{placements}. The figure shows that ACC values for five situations are all above 0.94 and are relatively stable, indicating the robustness of the proposed DST in detecting user stress under the different settings of sensor placement. Meanwhile, Figure \ref{placements} also reveals that the performance of DST tenderly increases with the growth of the angle, which is counter-intuitive. This phenomenon can be attributed to the decrease in respiration signal strength with the increase of the angle, which makes the heartbeat signal be less disturbed by the respiration signal~\cite{zhang2022can}.}

\subsubsection{\rev{Impact of Different Environmental Conditions}}
\rev{To show the robustness of DST in handling different environments, we have tested the effectiveness of DST under five environmental conditions, namely E1: classroom, E2: conference room, E3: living room, E4: laboratory, and E5: bedroom. 
Similarly to Sections \ref{sub:distance} and \ref{sub:angle}, three round of experiments are performed in every condition and $\frac{2}{3}$ collected physiological signal data of target user are treated as the training dataset while the remaining $\frac{1}{3}$ data are used as the test dataset. Experimental results displayed in Figure \ref{environments} show that ACC values of the proposed DST method under five environmental conditions are all above 0.9, which indicates the effectiveness of DST in coping with the stress detection problem under different environment settings. 
}


\subsubsection{Analysis of DASS-42 Stress Subscale}
In this section, the stress subscale of DASS-42 is used to verify the effectiveness of the stress-inducing protocol proposed in this paper. 
In Table \ref{tab:DASS} lists the mean ($M$) and standard deviation ($SD$) of the scores given by all target users on the 14 items of the stress subscale.
The $M$ value of each item is basically between 1.5 and 2.3, which indicates that all target users feel a certain degree of stress in the stress task. 
Table \ref{tab:DASS} also shows that most $SD$ values are between 0.6 and 1.1, and their variance is within a reasonable range. This means that the scores given by different target users are relatively consistent, which indicates the defined stress-inducing protocol can effectively induce stress for all target users. In DASS-42 Stress Subscale, a total score of all items which is larger than 10 points represents different levels of stress~\cite{parkitny2010depression}. 
According to $M\pm SD$ calculations, the total scores of all items for all target users are all above 10 points, which indicates that each target user has a moderate or higher level of stress when performing the stress task. 
All analysis results of DASS-42 stress subscale show that the stress task in the proposed stress-inducing protocol can effectively induce stress of target users. This is because both of the questions answered by the target users in a stress task and the target users' anticipation to the arrival of an unknown pulse triggered by wrongly answering a question cause stress.


\begin{table}[!t]
\begin{center}
\caption{The scores of the participants on each item about DASS-42 stress subscale}
\label{tab:DASS}
\begin{tabular}{|c|c|c|c|}
\hline
Item of question & M±SD & Item of question & M±SD \\ \hline
S1 & 2.1±0.8  & S8&   2.2±1.0      \\ \hline
S2 & 1.6±0.7  & S9&   1.8±0.9      \\ \hline
S3 & 1.9±0.9  & S10&  1.9±0.8       \\ \hline
S4 & 2.3±1.1  & S11&  2.1±0.9       \\ \hline
S5 & 1.5±0.6  & S12&  1.6±0.7       \\ \hline
S6 & 2.0±0.8  & S13&  1.8±0.8       \\ \hline
S7 & 1.7±0.8  & S14&  2.0±0.9       \\ \hline
\end{tabular}
\end{center}
\end{table}

\section{Conclusion and Future Work} \label{sec:discussion}
In this paper, we formally define the stress tracing problem and present DST, a novel deep network for tracing human stress. Compared with related stress detection \rev{methods}, DST well solve two vital challenges in stress detection domain, i.e., unfriendly \rev{collection of} physiological signals \rev{from} users and \rev{failure} to \rev{effectively utilize} multimodal physiological signals. 
\rev{Specifically}, unlike most related works that leverage wearable devices to collect physiological signals of users, DST proposes \rev{tracing} human stress based on physiological signals collected by a contactless UWB radar, which improves user experience. 
Then, a group of modules, namely\rev{, } \rev{the} signal extraction module (SEM) and multimodal fusion module (MFM) are carefully designed in DST to ensure \rev{that} the multimodal physiological signals of target users can be effectively extracted, fused, and utilized for tracing their stress states. 
Experimental results on three real-world datasets demonstrate that DST method delivers significant improvements over state-of-the-art baselines in tracing human stress. 
One limitation of this study is that we do not consider users' side information and his/her performance when  completing the task, which we will consider in the future to identify user stress levels at a finer granularity. 
For future directions, it is worth exploring how to deploy larger-scale
models such as large language models on resource-constrained
wearables~\cite{lin2024splitlora,fang2024automated,qiu2024ifvit} and how to collaboratively train models
while protecting user privacy~\cite{hu2024accelerating,lin2024adaptsfl,zhang2024satfed,zheng2023autofed,zhang2024fedac,lyu2023optimal,lin2023fedsn}.

\section*{Acknowledgements}
This work is supported by National Natural Science Foundation of China [Nos. 62067001, 62062008].










\bibliographystyle{unsrt}

\bibliography{cas-refs}

\end{document}